\documentclass[preprint,showpacs,12pt]{revtex4}
\usepackage{amsmath,graphicx,epsfig,epstopdf}
\usepackage[T1]{fontenc} 
\newcounter{saveeqn}

\begin{document}

\def\gsimeq{\,\,\raise0.14em\hbox{$>$}\kern-0.76em\lower0.28em\hbox {$\sim$}\,\,}
\def\lsimeq{\,\,\raise0.14em\hbox{$<$}\kern-0.76em\lower0.28em\hbox{$\sim$}\,\,}
\def\Msun{$M_{\odot}$}

\title{Large-scale deformed quasiparticle random-phase approximation calculations of the $\gamma$-ray strength function using the Gogny force}



\author{M. Martini$^{1}$, S. P\'eru$^{2}$, S. Hilaire$^{2}$, S. Goriely$^{3}$, F. Lechaftois$^2$}

\affiliation{$^{1}$ ESNT, CEA-Saclay, DSM, IRFU, Service de Physique Nucl\'eaire,  F-91191 Gif-sur-Yvette Cedex}

\affiliation{$^{2}$ CEA, DAM, DIF, F-91297 Arpajon, France} 

\affiliation{$^{3}$ Institut d'Astronomie et d'Astrophysique, Universit\'e Libre de Bruxelles, CP-226, 1050 Brussels, Belgium}

\date{\today}

\begin{abstract}

Valuable theoretical predictions of nuclear dipole excitations in the whole chart are of great interest for different nuclear applications, including in particular nuclear astrophysics.
Here we present large-scale calculations of the $E1$ $\gamma$-ray strength function obtained in the framework of the axially-symmetric deformed quasiparticle random phase approximation based on the finite-range Gogny force. 
This approach is applied to even-even nuclei, the strength function for odd nuclei being derived by interpolation.
The convergence with respect to the adopted number of harmonic oscillator shells  and the cut-off energy introduced in the 2-quasiparticle (2-$qp$) excitation space is analyzed. The calculations performed with two different Gogny interactions, namely D1S and D1M, are compared. A systematic energy shift of the $E1$ strength is found for D1M relative to D1S, leading to a lower energy centroid and a smaller energy-weighted sum rule for D1M. When comparing with experimental photoabsorption data, the Gogny-QRPA predictions are found to overestimate the giant dipole energy by typically $\sim$2 MeV. 

Despite the microscopic nature of our self-consistent
Hartree-Fock-Bogoliubov plus QRPA calculation, some phenomenological corrections need to be included to take into account the effects beyond the standard 2-$qp$ QRPA excitations and the coupling between the single-particle and low-lying collective phonon degrees of freedom. For this purpose, three prescriptions of folding procedure are considered and adjusted to reproduce experimental photoabsorption data at best. 
All of them are shown to lead to rather similar predictions of the $E1$ strength, both at low energies and for exotic neutron-rich nuclei. Predictions of $\gamma$-ray strength functions and Maxwellian-averaged neutron capture rates for the whole Sn isotopic chain are also discussed and compared with previous theoretical calculations. 

\end{abstract}




\maketitle

\section{Introduction}

About half of the nuclei with $A>60$ observed in nature are formed by the rapid neutron-capture process (r-process) occurring in explosive stellar events \cite{arnould07}. The r-process is expected in environments with high neutron density ($N_n > 10^{20}~{\rm cm^{-3}}$). Successive neutron captures proceed into neutron-rich regions well off the $\beta$-stability valley forming exotic nuclei that cannot be produced and therefore studied in the laboratory. 
When the temperature or the neutron density required for the r-process are low enough to break the (n,$\gamma$)-($\gamma$,n) equilibrium, the r-abundance distribution depends directly on the neutron capture rates of the so-produced exotic neutron-rich nuclei \cite{go98}. The neutron capture rates are commonly evaluated within the framework of the statistical model of Hauser-Feshbach, although the direct capture contribution play an important role for very exotic nuclei \cite{xu14}. The fundamental assumption of the Hauser-Feshbach model is that the capture go through the intermediary formation of a compound nucleus in thermodynamic equilibrium. In this approach, the Maxwellian-averaged (n,$\gamma$) rate at temperatures of relevance in r-process environments strongly depends on the electromagnetic interaction, i.e the photon de-excitation probability. The well known challenge of understanding the r-process abundances thus requires reliable extrapolations of the $E1$-strength function out towards the neutron-drip line. 

Large scale calculations of $E1$ $\gamma$-ray strength functions are usually performed on the basis of the  phenomenological Lorentzian model \cite{ripl3}. The reliability of the $\gamma$-ray strength predictions can thus be greatly improved by the use of microscopic models. 
Indeed, provided satisfactory reproduction of available experimental data, the more microscopic the underlying theory, the greater the confidence in the extrapolations out towards the experimentally unreachable regions. 
Microscopic approaches are rarely used for practical applications. First, the time cost is often prohibitive for large scale calculations. 
Second, the fine tuning required to reproduce accurately a large experimental data set is very delicate, in addition to be time consuming. 
A prominent exception is represented by Refs. \cite{Gor02,Gor04} where a complete set of $\gamma$-ray strength functions was derived from mean field plus quasiparticle random phase approximation (QRPA) calculations. In Refs. \cite{Gor02,Gor04}, zero-range Skyrme forces were considered and phenomenological corrections applied to properly describe the splitting of the giant dipole resonance (GDR) in deformed nuclei as well as the damping of the collective motion.

The present study aims to go beyond the former approximation providing axially-symmetric-deformed QRPA approach based on Hartree-Fock-Bogoliubov (HFB) calculations using the finite-range Gogny interaction in a fully consistent way. Contrary to the calculations in which radial wave functions are discretized on a mesh, single-particle wave functions are here expanded on an optimized harmonic oscillator (HO) basis. 
The present approach is specially suited for open-shell nuclei, where pairing correlations are included without any additional parameters. 
Another asset is an adequate treatment of the deformation at each step of the calculation. 
More precisely the intrinsic deformation of the nucleus ground state is predicted by the HFB calculations as the minimum of the potential energy surface. Then the QRPA phonons are oscillations around this minimum, spherical or not. 
This treatment significantly improves the description of the nuclear structure property, hence the $\gamma$-ray strength function predictions. 
It was first applied to study giant resonances in Si and Mg isotopes \cite{PG08}. Dipole excitations in Ne isotopes and $N=16$ isotones \cite{Mar11} as well as electromagnetic excitations of the heavy deformed $^{238}$U \cite{Per11} have been obtained with an optimized version of the numerical code, opening the way to large scale calculations. 
This powerful tool has been generalized to treat also charge exchange excitations which are relevant for the $\beta$-decay of experimentally inaccessible nuclei \cite{Martini:2014ura}. 
In the present paper, we test the predictive power of the aforementioned approach by applying it to a large set of even-even nuclei 
in order to compare with available experimental data on photoabsorption~\cite{ripl3}, the $E1$ $\gamma$-strength function of odd nuclei being obtained through an interpolation procedure involving neighbouring even-even nuclei.

The paper is organized as follows. 
In Sec.~\ref{sec_mod}, the axially-symmetric-deformed HFB+QRPA formalism is described in its standard form and possible extensions are sketched. In the same spirit as Ref.~\cite{DechargeNPA83}, the impact of the size of the finite HO basis including cut-off effects are analyzed in Sec.~\ref{sec_sens} adding discussion on the choice of the interaction parameter sets.
This convergence analysis sets a protocol for large scale calculations whose results are presented 
in Sect.~\ref{sec_exp}. First, the impact on deformation of the $\gamma$-strength function is illustrated. Second, the comparison with photo absorption data ~\cite{ripl3} is shown. Third, models are introduced to obtain the continuous strength functions starting from the discrete QRPA strength distributions $B(E1)$ using or not  microscopic input. The final $E1$ strength functions as well as the corresponding Hauser-Feshbach astrophysical reaction rates are finally estimated for a large set of exotic neutron-rich nuclei and compared with other predictions. Conclusions are drawn in Sect.~\ref{sec_conc}. 

\section{The theoretical model}
\label{sec_mod}

\subsection{Standard HFB+QRPA approach} \label{formalism}

We summarize here the formalism of the consistent QRPA approach based on axially-symmetric-deformed HFB equations solved in a finite HO basis. For more details we refer the reader to Refs. \cite{PG08,Per14}. With rotational invariance along the $Z$ axis and time reversal symmetry, the eigenstates of the HO basis are fully identified by three spatial quantum numbers $m$, $n_{\bot}$ and $n_z$ plus the spinor $\sigma$. For each HO state, the projection of total angular momentum onto the symmetry axis is $k=m+\sigma_z$ and the parity is $\pi = (-)^{|m| + n_z}$. 
Imposing here a same HO pulsation in $Z$ and perpendicular directions, the energy quantum number of each state is given by $N= |m|+ 2 n_{\bot} + n_z $. 
The upper limit $N_0\geq N$ of a finite basis gives the number of involved major shells $N_{sh}=N_0+1$, namely the size of the basis. 
Usually in HFB $N_0$ is chosen according to the rule that the number of HO states is 8 times the maximum of proton or neutron occupied states. In the present calculation the HO pulsation is adjusted for each nucleus at each deformation and for each number of major shells in order to minimize the HFB binding energy. Solving the HFB equations in HO basis leads to the diagonalisation of a Hamiltonian matrix: eigenvalues and eigenvectors are respectively Bogoliubov quasi-particle ($qp$) excitation energies and $u$ and $v$ components of the Bogoliubov transformation. As a consequence the positive energy continuum is discretized. 
The first order excitations for even-even nuclei are given by two-quasiparticle (2-$qp$) excitations. 
QRPA phonons are linear combinations of these 2-$qp$ excitations.  
According to the symmetries imposed, the projection $K$ of the angular momentum $J$ on the symmetry axis and the parity $\Pi$ are good quantum numbers for the phonons. Consequently, QRPA calculations can be performed separately for each $K^{\Pi}$ set. In this context, phonons are characterized by the excitation operator
\begin{equation}\label{thetaplus}
\theta^+_{n,K^{\Pi}}=\sum_{ij} X^{ij}_{n,K^{\Pi}} \eta^+_i \eta^+_j-(-)^K Y^{ij}_{n,K^{\Pi}} \eta_j \eta_i,
\end{equation}
where $\eta^+$ and $\eta$ are the quasi-particle operators, related to the HO particle operators $c^+$ and $c$ through the $u$ and $v$ Bogoliubov transformation matrices:
\begin{equation}\label{bogo_transf}
\eta^+_i=u_{i \alpha} c^+_\alpha-v_{i \alpha} c_\alpha.
\end{equation}
Here and in the following, repeated indices are implicitly summed over; latin and greek letters denote quasiparticle and harmonic oscillator states, respectively. Conservation of symmetries imposes  $k^{\pi_i}_i=k^{\pi_\alpha}_\alpha$ in Eq.(\ref{bogo_transf}) and that the condition $K=k_i+k_j$ in Eq.(\ref{thetaplus}) and $\Pi =\pi_i  \pi_j$ are fulfilled. 
According to the occupation probabilities, the quasi-particle creation operator $\eta_i^+$ is associated to the creation operator of particle type $c_p^+$ when $u_i^2 >> v_i^2$, and it corresponds to annihilation operator of hole type $c_h$ if $u_i^2<< v_i^2$. 
In principle QRPA calculation can be performed without any cut-off in energy of the 2-$qp$ states neither in occupation probabilities ($v^2$) of single quasiparticle states. 
The amplitudes $X$ and $Y$ of Eq. (\ref{thetaplus}) are solutions of the well-known QRPA matrix equation \cite{book_ring_schuck}
\begin{equation}\label{equaref}
\left(\begin{array}{cc} { A}& { B}\\{ B}&{ A} \end{array}  \right)
 \left(\begin{array}{c}{X_{n,K^{\Pi}}}\\ {Y_{n,K^{\Pi}}}\end{array}\right)
= \omega_{n,K^{\Pi}} \left(\begin{array}{c}X_{n,K^{\Pi}}\\-Y_{n,K^{\Pi}}\end{array}\right),
\end{equation}
where  $\omega_{n,K}$ are the energies of the QRPA excited  states. 
To ensure consistency, the same interaction (parameter set of the Gogny force) is used to calculate the $A$ and $B$ matrix elements and the underlined HFB mean field \cite{Per14}. We consider in the present study the D1S and D1M forces only, whose properties and parameters are summarized in appendix of Ref. \cite{Per14}. 
Once the QRPA matrix is diagonalised, the $X$ and $Y$ amplitudes allow to calculate the strength for each electromagnetic mode. 
Here we focus on the dipole (multipolarity $\lambda=1$, parity $\Pi=(-)$) mode. 
The isovector dipole excitation operator is
\begin{equation}
\hat{Q}_{1,0}= \sum_{i}^{Z}r_i\mathcal{Y}_{1,0}(\theta_i,\phi_i) -\sum_{i}^{N}r_i\mathcal{Y}_{1,0}(\theta_i,\phi_i),
\end{equation}
where 
$\mathcal{Y}_{1,0}(\theta_i,\phi_i)$ is a spherical harmonic. 
This isovector operator does not excite too much the center of mass motion whose nature is essentially isoscalar 
(more informations on center of mass spurious states are given in Sec.~\ref{sec_bases}).  

The total dipole distribution $B(E1)$ (in $e^2$fm$^2$) is obtained by summing the contributions of $K^\Pi=0^-$ and twice that of $K^\Pi=1^-$, the $K^\Pi=-1^-$ solution being equal to the $K^\Pi=1^-$ one through the conservation of time reversal symmetry. We remind that in the spherical symmetry case, the $K=0$ and $|K|= 1$ states are degenerate. In deformed nuclei, the dipole strength splits up into two components corresponding to two different angular momentum projections $K$. 

\subsection{Beyond 2-$qp$ excitations} \label{formalism_4qp}
The well established formalism described above takes into account only 2-$qp$ excitations. 
 Two extensions of the standard (Q)RPA approach have been developed in the past by different groups: the so-called second RPA 
\cite{Yannouleas:1983kp,Yannouleas:1987zz} and the particle-vibration coupling \cite{Bertsch:1983zz} or quasiparticle-phonon model \cite{Sol76,Sol78,Tso08}. 
In the second RPA calculations the centroid energy of the response function is shifted by a few MeV to lower energies with respect to the standard QRPA values, as shown and discussed in Refs.~\cite{pap09,gam11,gam12,Gambacurta:2015pva}. The interaction between the single-particle and low-lying collective phonon degrees of freedom also shifts the $E1$ strength to lower energies \cite{Colo:2001fz,kame04,Sarchi:2004pf,avde11,acha15}. 
A fragmentation and a broadening of the response also appears within these approaches. 
To include these effects, the discrete $B(E1)$ distribution is usually folded by a Lorentzian function 
\begin{equation}
L(E,\omega)=\frac{1}{\pi}\frac{\Gamma E^2}{[E^2-(\omega-\Delta)^2]^2+\Gamma^2E^2} \quad \label{broad},
\end{equation}
leading to a continuum result for the $E1$ $\gamma$-ray strength function $S_{E1}(E)$ (in $e^2$fm$^2$ MeV$^{-1}$):
\begin{equation}
S_{E1}(E)=\sum_{n} L(E,\omega_n)B(E1)(\omega_n).
\label{eq_folding}
\end{equation}
In Eq.~(\ref{broad}), $\Gamma$ is the width at half maximum and $\Delta$ allows for an energy shift. 
These quantities could be adjusted on experimental data or should be obtained by the aforementioned beyond (Q)RPA formalism. 
The generalization of QRPA would include 4- and 6-$qp$ excitations up to infinity. Increasing the $n$ order of $n$-$qp$ excitations 
provides a shifted spectrum closer and closer to the continuum one, especially in the giant resonance region. 
The energy shift $\Delta$ is expected to be related to the density of $n$-$qp$ states, hence, in a first approximation, to the density of 4-$qp$ states. 
In the present study this quantity has been calculated for dipole excitations. 
Note that for each $K$ value, the number $n^K_{4qp}$ takes into account only the configurations involved in the dipole excitation. 
Quantum numbers and occupation probabilities drive the selections of the relevant 4-$qp$ states conserving isospin ($\tau$) and particle number. Once the Wick theorem is applied to a 4-$qp$ excitation operator $\eta^+_{i}\eta^+_{j}\eta^+_{k}\eta^+_{l}$, only the fully contracted terms ${\eta^+_{i}}^\bullet {\eta^+_{j}}^\bullet~{\eta^+_{k}}^{\bullet\bullet}{\eta^+_{l}}^{\bullet\bullet}$, $(-){\eta^+_{i}}^\bullet {\eta^+_{k}}^\bullet~{\eta^+_{j}}^{\bullet\bullet}{\eta^+_{l}}^{\bullet\bullet}$ and ${\eta^+_{i}}^\bullet {\eta^+_{l}}^\bullet~{\eta^+_{j}}^{\bullet\bullet}{\eta^+_{k}}^{\bullet\bullet}$ need to be considered. 
If at least one of the three combinations contains an excitation of type 
${c^+_{p\tau}}^\bullet{c_{h\tau}}^\bullet~{c^+_{p'\tau'}}^{\bullet\bullet}{c_{h'\tau'}}^{\bullet\bullet}$, this 4-$qp$ excitation could contribute to the count of $n^K_{4qp}$. Additional constraints arise from the total parity $\Pi$ and $K$ conservation, $K= k_i+k_j+k_k+k_l=k_p+k_{p'}-k_h-k_{h'}$. These quantities should be equal to those of the QRPA calculation in the 2-$qp$ basis. According to angular momentum algebra the possible $k$ values $(k_p-k_h,k_{p'}-k_{h'})$ should be $(0,0)$, $(1,-1)$ or $(-1,1)$ for $K=0$, and $(1,0)$, $(0,1)$, $(-1,2)$ or $(2,-1)$ for $K=1$. 
The improvement of the folding procedure by the insertion of the $n^K_{4qp}$ quantity will be discussed in Sect.~\ref{sect_phon}.

\section{Sensitivity analysis}
\label{sec_sens}
For practical large-scale calculations, choices have to be made in order to limit the computational cost. 
As previously mentioned, the HFB+QRPA results remain sensitive to the choices made for the calculation. 
These choices include the number of HO shells used, the energy cut-off  $\varepsilon_c$ accounted for on the 2-$qp$ state energies, 
and the set of parameters of the Gogny interaction. We discuss below the impact of these effects on the calculated $E1$ strength function, with an additional view on the possibility to reduce the computational time without reducing the predictive power of the calculations. 

\subsection{Convergence with respect to the size of the basis} \label{sec_bases}

A key aspect to which the QRPA predictions are sensitive is the number of HO shells, $N_{sh}$, included in the HFB+QRPA calculation. In order to investigate the impact of $N_{sh}$ on the $\gamma$-ray strength function and to verify the convergence, we have performed, using the D1M Gogny interaction \cite{Gor09}, QRPA calculations for several odd values of $N_{sh}$ without any energy cut-off on the 2-$qp$ states energies. Thus, in such a situation, whatever the size of the basis is, the center of mass spurious state is decoupled to the physical excitation spectrum and easily identified, as shown for example in Refs. \cite{Mar11,Per11,Per14}. 
Once the spurious state is identified, it is removed from the QRPA spectra.   

\begin{figure}[h]
\begin{center}
\includegraphics[trim=1cm 15.5cm 2cm 1cm, clip=true,scale=0.55]{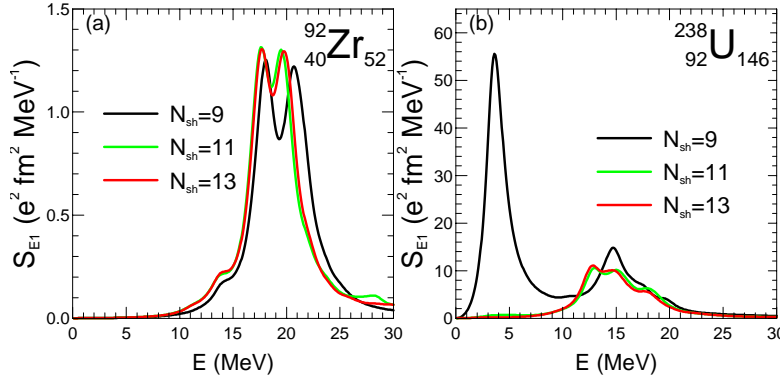}  
\caption{QRPA $E1$ strength $S_{E1}$ calculated with the D1M Gogny interaction as a function of the excitation energy for different numbers of HO shells without any energy cut-off on the 2-$qp$ state energies.}
\label{base1}
\end{center}
\end{figure}

The results are plotted in Fig.~\ref{base1} for two nuclei, namely the spherical $^{92}$Zr nucleus and the well deformed actinide $^{238}$U using the broadening procedure of Eq.~(\ref{broad}) with a width $\Gamma=2$ MeV and no shift ($\Delta=0$).

As can be observed, the QRPA strengths are shifted towards lower energies when $N_{sh}$ is increased. 
In Fig.~\ref{base1}(b), we can even observe for $N_{sh}=9$ a large strength below 5 MeV which disappears for larger bases. 
Since the QRPA calculations without cut-off are completely consistent, no spurious states are involved in this $N_{sh}=9$ low energy strength 
which is rather unphysical and illustrates that using a too low number of HO shells is inappropriate for heavy nuclei.  
Another important feature observed in this plot is the convergence of the predictions with increasing $N_{sh}$. It is well illustrated by the fact that for $N_{sh}=11$ and $13$, QRPA strengths are very close to each other, as compared to $N_{sh}=9$.
\begin{table}
\begin{center}
\caption {Average computation time for a $K^{\pi}=0^-$ of one nucleus for several energy cut-off  and basis size combinations using 1024 cpus.}
\begin{tabular}{|c|c|c|c|c|}
\hline
          $N_{sh}$  &   No cut  &   $\varepsilon_c=100$ MeV &   $\varepsilon_c=60$ MeV &  $\varepsilon_c=30$ MeV  \\
\hline
       9  & 5 min    & 5 min   & 4 min   & 38 s\\ 
      11  & 2 h   & 2 h  & 1h   &  5 min \\ 
      13  & 42 h  & 26 h & 6 h  & 30 min \\
      15  & 21 d  & 8 d  & 30 h &  2 h \\
      17  & 286 d & 63 d & 7 d  &  8 h \\
\hline
\end{tabular}
\label{time}
\end{center}
\end{table}

\begin{figure}[h]
\begin{center}
\includegraphics[trim=0cm 15cm 1cm 1cm, clip=true, scale=0.75]{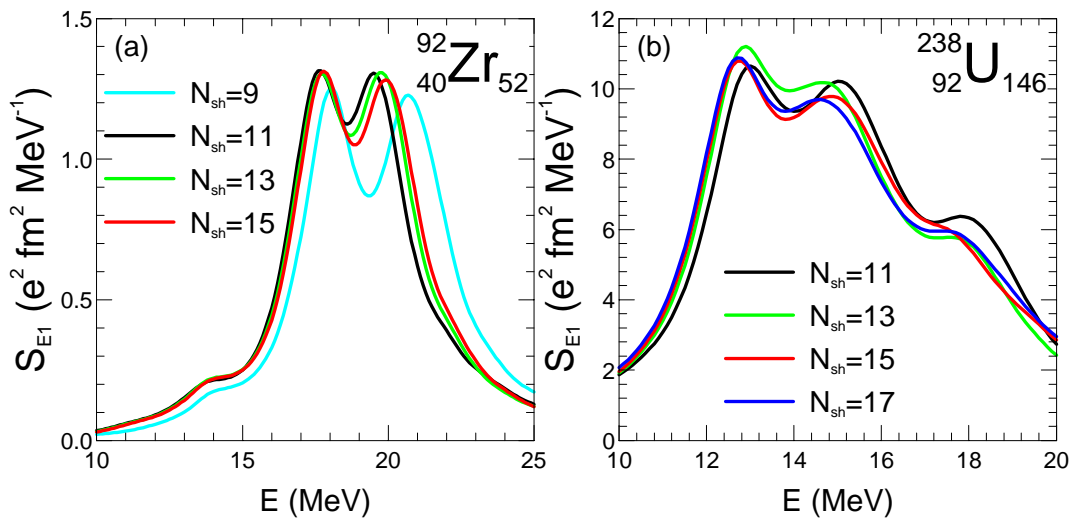}  
\caption{(Color Online) QRPA $E1$ strength $S_{E1}$ calculated with the D1M Gogny interaction for different numbers of HO shells for (a) $^{92}$Zr with an energy cut-off on the 2-$qp$ states energies of $\varepsilon_c=120$~MeV and (b) for $^{238}$U with $\varepsilon_c=60$~MeV.}
\label{base2}
\end{center}
\end{figure}

Considering the known overestimation of the order of 2 MeV of the energy position of the QRPA $E1$ strength function in $^{238}$U \cite{Per11}, it is important to be able to disentangle between model limitations and lack of convergence. We therefore display in Fig.~\ref{base2} a zoom of the energy peak region in order to better estimate the peak position as a function of $N_{sh}$, including also higher $N_{sh}$ values than in Fig.~\ref{base1}, and doing so, to confirm the convergence observed in Fig.~\ref{base1}. However, we now use an energy cut-off $\varepsilon_c=120$~MeV for  $^{92}$Zr and $60$ MeV for $^{238}$U, independently of the adopted $N_{sh}$ value. 

The choice of a constant cut-off avoids unreasonable computation time for $N_{sh}=15$ and $17$ and prevents interference between the impact of energy cut-off and basis size. As can be observed in Table \ref{time}, the computation time, without any energy cut-off is clearly unreasonable as soon as more than $15$ shells are used. For $N_{sh} \geq 15$, the only way to reduce the computation time to an acceptable limit consists in introducing an energy cut-off which, for $N_{sh}=17$ is at most 60 MeV if we want to remain within a feasible range. 
Figure~\ref{base2} confirms 
the fact that with $N_{sh}=13$ for $^{92}$Zr and 
$N_{sh}=15$ for $^{238}$U one is very close to converged values as far as the peak positions are concerned. Quantitatively speaking, the change in the peak energies is of the order of 1 MeV, 200 keV and 100 keV between $N_{sh}=9$ and 11, 11 and 13 and 13 and 15 for $^{92}$Zr and of the same magnitude for $^{238}$U between $N_{sh}=11$ and 13, 13 and 15 and 15 and 17, respectively. Comparing Fig.~\ref{base1}(b) and \ref{base2}(b), one also observes that a cut-off of 60 MeV  provides similar results as a calculation without any cut-off for $^{238}$U.

\subsection{Role of the 2-$qp$ excitation energy cut-off}

As shown in Table \ref{time} and already discussed above, the energy cut-off is a way to get a compromise between the computational time and convergence of the calculation. This compromise begins to be interesting for $N_{sh} \geq 13$ and is clearly necessary for $N_{sh} \geq 15$. We have therefore studied the convergence of the calculation as a function of the energy cut-off. We have selected both spherical and deformed nuclei as well as light, medium-mass and heavy nuclei in order to see if the energy cut-off ensuring an acceptable convergence depends on the number of shells or in other words on the nucleus mass.  
\begin{figure}
\begin{center}
\includegraphics[trim=0.7cm 2cm 0cm 0cm, clip=true, scale=0.58]{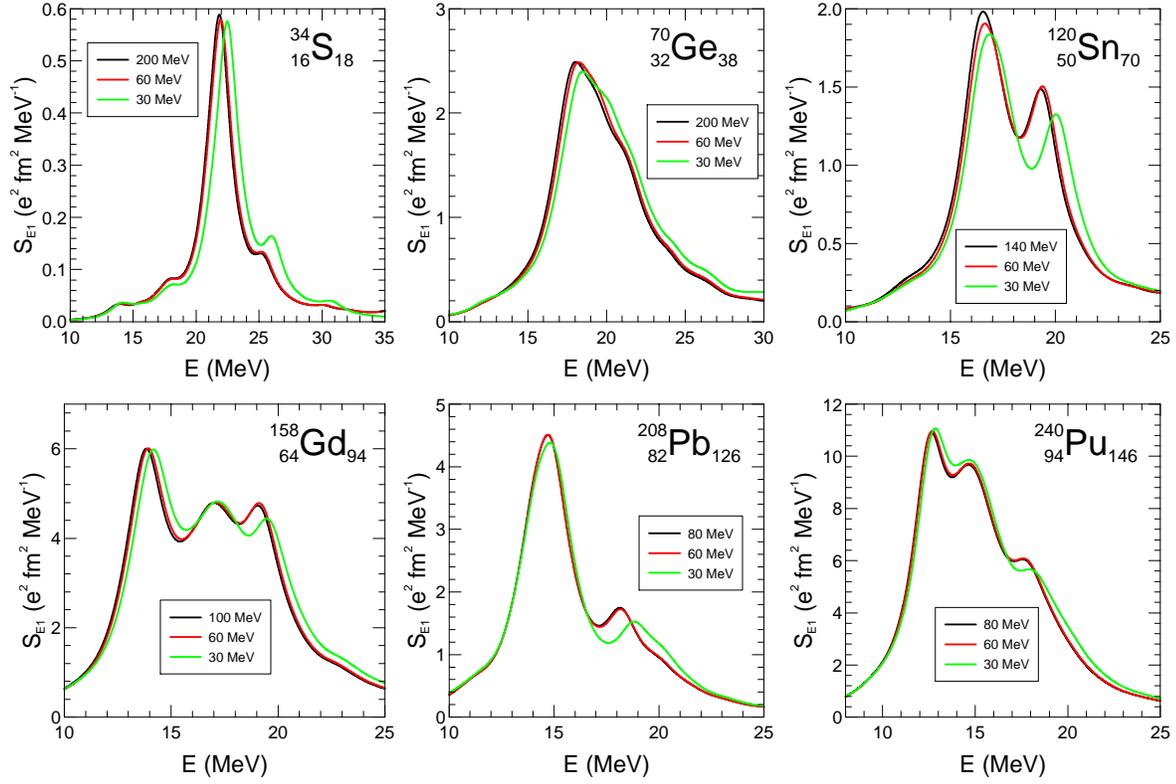}  
\caption{QRPA $E1$ strength $S_{E1}$ calculated with the D1M Gogny interaction for different cut-off energies for an arbitrary sample of nuclei. The number of HO shells chosen for each 
nucleus ensures convergence with respect to the conclusions of Sect. \ref{sec_bases} and is summarized in Table \ref{shells}.}
\label{cuts}
\end{center}
\end{figure}
Typical results are plotted in Fig.~\ref{cuts} for several nuclei both deformed and spherical as a function of three energy cut-off values using again the broadening procedure of Eq.~(\ref{eq_folding}) with a width $\Gamma=2$ MeV (and no shift, i.e. $\Delta=0$). As can be observed, independently of the basis size ($N_{sh}$ increases with the nucleus mass), an energy cut-off $\varepsilon_c=60$~MeV ensures a satisfactory convergence of the $E1$ strength functions as far as the peak energy is concerned, while a $\varepsilon_c=30$~MeV  is clearly not sufficient although it already provides the main part of the strength when compared to a higher energy cut-off. This feature is confirmed in Fig.~\ref{cuts2} where the energy shift of the first and second energy peaks is plotted as a function of the nuclear mass for a large set of nuclei covering the whole mass range of our study. This shift is defined as the difference between the peak energies for a given energy cut-off with respect to the peak energy without any cut-off or with a high enough energy cut-off (80 MeV in the actinide region for instance) to consider that convergence has been reached. It is worth mentioning here that for some nuclei the size of the basis is already the one which will be finally adopted in the present study (see Table \ref{shells}). For these nuclei no major impact of the basis size is observed on the peak energy shifts. 

\begin{figure}
\begin{center}
\includegraphics[trim=0cm 13.5cm 0cm 3cm, clip=true, scale=0.8]{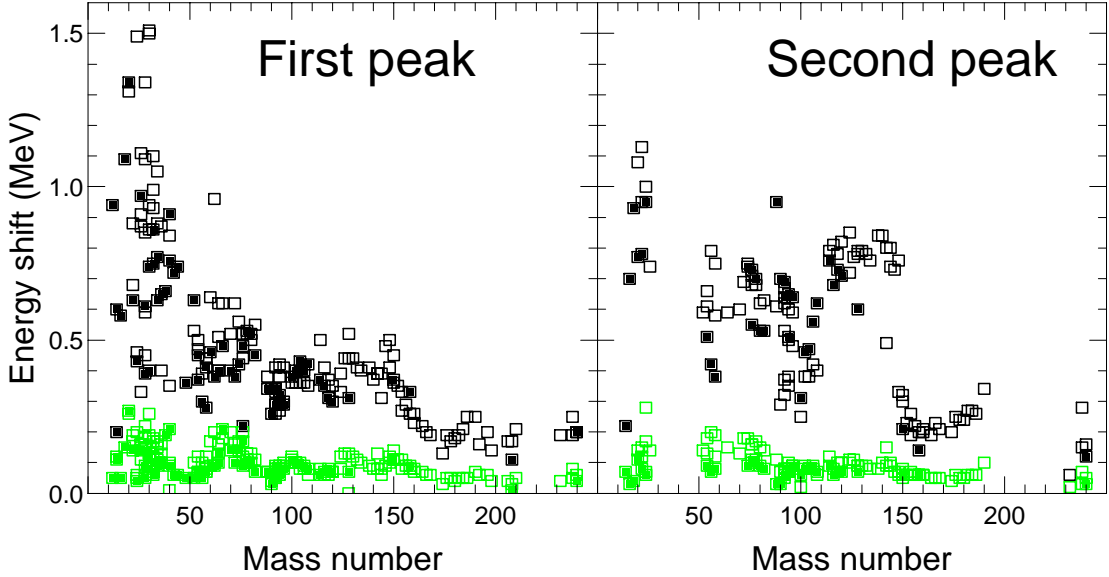}  
\caption{Energy shifts of the two main peaks of the $B(E1)$ distribution predicted with D1M as a function of the atomic mass for two different values of the cut-off energy (see text for details). Black (resp. green) squares correspond to a $\varepsilon_c=30$~MeV (resp. 60 MeV). 
The full squares indicate calculations for which $N_{sh}$ values are compatible with Table \ref{shells}.}
\label{cuts2}
\end{center}
\end{figure}

Considering the computational price to pay to include 2-$qp$ states with energies higher than 60 MeV with respect to the gain in the accuracy, it is worth performing QRPA calculations with a cut-off of 60 MeV rather than without any energy cut-off. In this case, the difference in the peaks positions is of the order of 100 to 200 keV which is acceptable if one keeps in mind the 2 MeV difference between experiment and theory observed in Ref. \cite{Per11} (see Sect.~\ref{sec_exp}). On the contrary, a 30 MeV cut-off gives rise to unacceptable shifts in the peak positions, with respect to the converged value, of the order of 0.5 to 1~MeV. 
The energy shift of the first peak of the 30 MeV cut-off decreases with the mass number following a low $A^{-\alpha}$. 
Recalling that the empirical mass dependence for the excitation energy of the isovector GDR has been found \cite{Berman:1975tt} 
to lie between $A^{-1/3}$ and $A^{-1/6}$, the GDR energy of light nuclei is relatively larger, hence closer to the cut-off, in comparison with those found in heavy nuclei. 
This explains the large shift between full and low energy cut-off calculations observed in Fig.~\ref{cuts2} for light nuclei. 

\subsection{Impact of the interaction parameters}

\begin{figure}
\begin{center}
\includegraphics[trim=1cm 1cm 1cm 1cm, clip=true, scale=0.78]{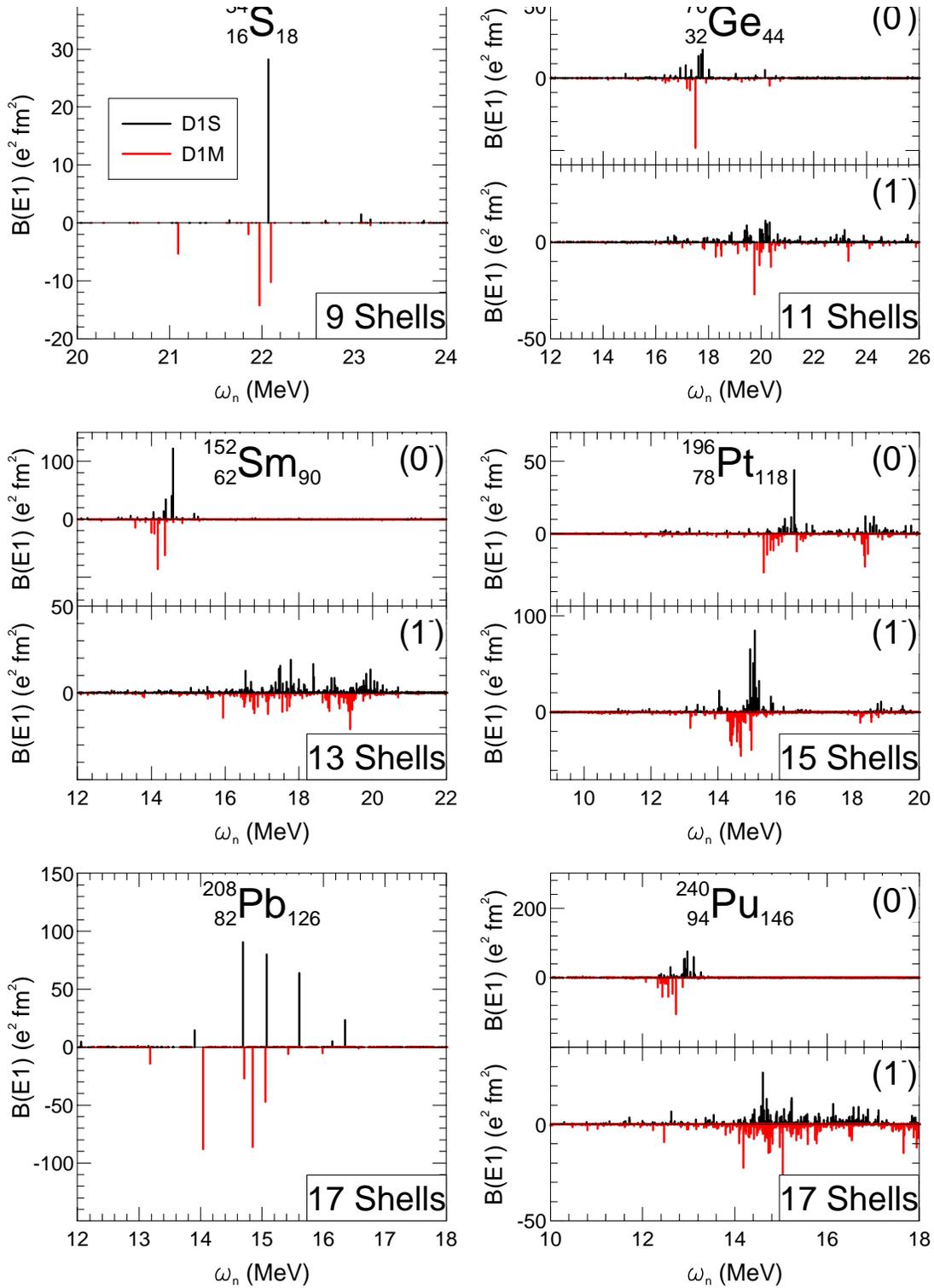}  
\caption{QRPA $B(E1)$ distributions in the GDR region for a reduced set of nuclei calculated with the D1S and D1M Gogny interactions for identical energy cut-off and basis size. For deformed nuclei ($^{76}$Ge, $^{152}$Sm, $^{196}$Pt and $^{240}$Pu) $K^\Pi=0^-$ and $|K|^\Pi=1^-$ contributions are plotted separately. The D1M strengths are multiplied by -1 for clarity.}
\label{inter}
\end{center}
\end{figure}

The predictions obtained with two different parameter sets of the Gogny interaction, namely D1S and D1M, are now compared, using the 2-$qp$ energy cut-off of 60 MeV as justified above and the same basis size for each nucleus. We first consider a reduced set of nuclei, both spherical and deformed and plot in Fig.~\ref{inter} the $B(E1)$ distribution obtained with both D1S and D1M interactions. Here we present the $B(E1)$ discrete distribution instead of the folded $S_{E1}$ in order to investigate possible differences in the strength fragmentation and peaks position due to the interaction itself.  

As can be observed, both interactions provide comparable results even though D1M seems to provide, systematically, slightly lower energy peaks than D1S, independently of $K^\Pi$. Focussing on deformed nuclei, one observes that for both interactions, the energy split between $K=0^-$ and $|K|=1^-$ states, related to the intrinsic deformation of the HFB ground state, follows an opposite hierarchy for prolate ($^{152}$Sm) and oblate ($^{196}$Pt) shapes, a result similar to what has already been discussed in Ref. \cite{PG08} for a different set of nuclei and also found systematically with a Skyrme interaction \cite{SL14}.

\begin{figure}
\begin{center}
\includegraphics[trim=1cm 2cm 2cm 2.5cm, clip=true, scale=0.5]{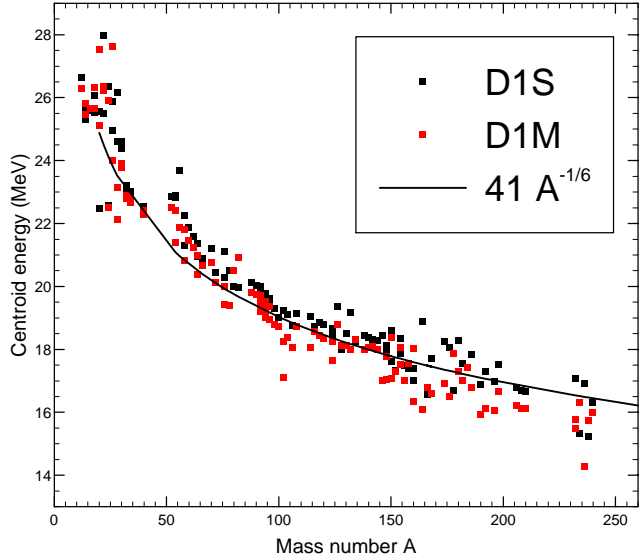}  
\caption{Energies of the centroid of the QRPA $B(E1)$ distributions calculated with D1S and D1M Gogny interactions for all the nuclei considered in this work. The full line represents the fit of the points.}
\label{centroid}
\end{center}
\end{figure}

To further investigate the tendency of D1M to yield lower centroid energies than D1S, we extend our study to a larger set of nuclei. However, to limit  computation time, lower $N_{sh}$ values than those of Table \ref{shells} are used. We plot in Fig. \ref{centroid} the centroid energies defined for each nucleus by
\begin{equation}
C=\frac{\sum_n \omega_n B(E1)(\omega_n)}{\sum_n B(E1)(\omega_n)} \quad ,
\end{equation}
where $\omega_n$ are the discrete energies of the QRPA states solution of Eq. (\ref{equaref}) and $B(E1)(\omega_n)$ the corresponding $B(E1)$ distributions.

As can be observed, the tendency to obtain lower energy peaks with D1M (Fig. \ref{inter}) is confirmed and is a rather general feature of our results. 
This shift is of the order of a few hundred keV, along the whole mass range. Moreover, we clearly observe that both interactions provide similar
qualitative behaviour with respect to the mass dependence of the energy peak positions. This behavior can be fitted for example by the 
law $E=41 A^{-1/6}$ MeV, a nuclear mass dependence compatible with the empirical \cite{Berman:1975tt} and theoretical \cite{VanIsacker:1992zz} behavior of the GDR energies.

\begin{figure}
\begin{center}
\includegraphics[trim=2cm 6cm 13cm 5cm, clip=true, scale=0.8]{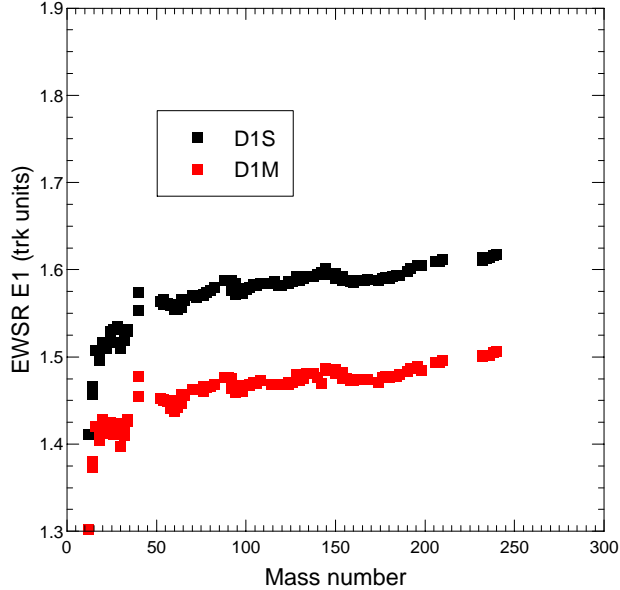}  
\caption{$E1$ EWSR in TRK units (14.8 $\frac{NZ}{A}$ e$^{2}$ fm$^{2}$ MeV) calculated in QRPA with D1S and D1M Gogny interactions for all the nuclei considered in this work.}
\label{fig_ewsr}
\end{center}
\end{figure}

The $E1$ energy-weighted sum rules (EWSR) expressed in Thomas-Reiche-Kuhn (TRK) units are plotted in Fig.~\ref{fig_ewsr} as a function of the mass number. 
The two parameter sets D1M and D1S of the Gogny interaction provide similar trends, also close to the one shown in Ref.~\cite{DechargeNPA83} obtained for a set of closed-shell nuclei with D1. The few hundred keV shift of the  centroid energy explains the decrease of the D1M EWSR with respect to D1S.

\subsection{Practical choices for large scale calculation}

Both the 2-$qp$ energy cut-off and the number of major shells are optimized to reach convergence in the predictions. 
For the study of giant resonances, the computational price to pay to obtain full convergence using the highest $N_{sh}$ value together without any energy cut-off is clearly not worth compared to the results obtained with an appropriate energy cut-off and a reasonable $N_{sh}$ value.

\begin{figure}[h]
\begin{center}
\includegraphics[trim=1cm 7cm 0cm 2.2cm, clip=true, scale=0.6]{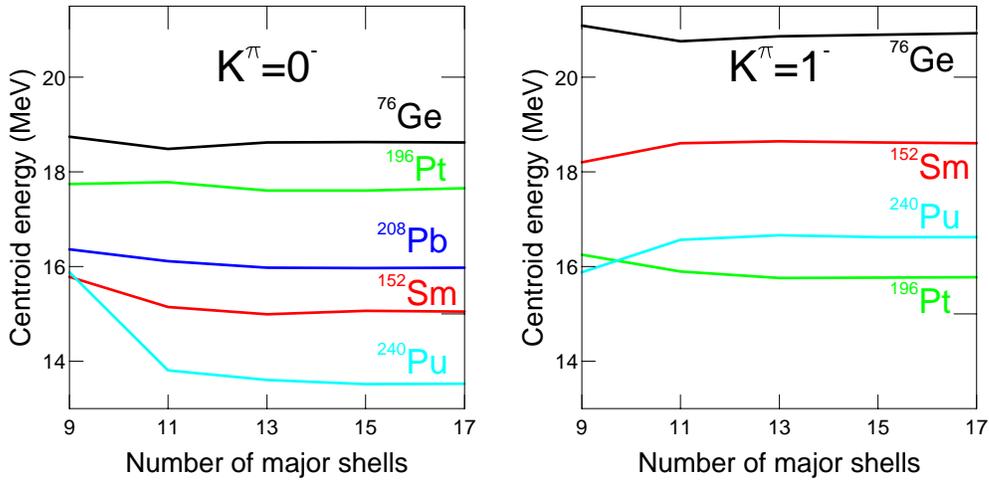}  
\caption{Energy centroid of the QRPA $B(E1)$ distributions calculated with D1M as a function of the number of shells using a global $60$ MeV energy cut-off for a sample of oblate ($^{198}$Pt), spherical ($^{208}$Pb) and prolate ($^{76}$Ge, $^{152}$Sm, $^{240}$Pu) nuclei.}
\label{def2}
\end{center}
\end{figure}

This is illustrated in Fig.~\ref{def2}, where the energies of the centroid of the predicted QRPA strengths are plotted as a function of the number of shells used in the calculation for a few nuclei with a fixed energy cut-off of 60 MeV. As can be observed, the centroid energies converge with increasing $N_{sh}$ values, and are only significantly different from that obtained with a highest $N_{sh}$ when the number of shells is clearly unreasonably low given the mass of the nucleus considered (9 shells for $^{152}$Sm, $^{196}$Pt or $^{240}$Pu for instance). We have therefore decided to adopt for all the calculations that will be discussed from now on a cut-off energy of 60 MeV and the $N_{sh}$ values, as summarized in Table \ref{shells}. 
\begin{table}
\begin{center}
\caption {Adopted number of HO shells  as a function of the nucleus}
\begin{tabular}{|c|c|c|c|c|c|}
\hline
          max$(N,Z)$  &   $\leq 20$  & $\leq 40$  & $\leq 70$ & $\leq 116$ & $> 116$  \\
\hline
            $N_{sh}$  &       9      &    11      &      13   &    15      &    17       \\ 
\hline
\end{tabular}
\label{shells}
\end{center}
\end{table}
It is worth mentioning that $N_{sh}=17$ have been used for a rather limited number of nuclei.
For more systematic studies in the heavy mass regions it would be possible to decrease this value to $N_{sh}=15$ without loosing too much 
in terms of quality of the predictions.

\section{Results and comparisons with data}
\label{sec_exp}

\subsection{Impact of the deformation} \label{sbsec_def}

\begin{figure}
\begin{center}
\includegraphics[trim=0cm 16cm 1cm 0.5cm, clip=true, scale=0.7]{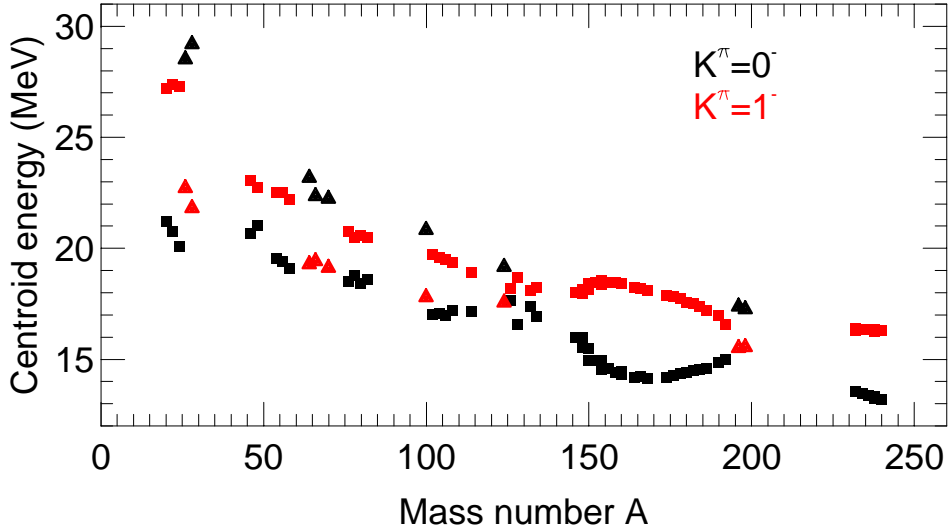}  
\caption{Centroid energies of the $K^{\pi}=0^-$ and $1^-$ components for all the deformed nuclei considered in this work. The squares correspond to prolate nuclei and the triangles to oblate ones.}
\label{defsys}
\end{center}
\end{figure}

As already discussed in Refs. \cite{PG08, Yoshida:2008rw, Kleinig:2008gq, Arteaga:2009mb, Arteaga:2008ej, Losa:2010bm}, 
for deformed nuclei the GDR splits in two energy components due to the split between 
$K=0$ and $K=\pm 1$ states. 
It has been noticed  \cite{PG08} that this split follows an opposite hierarchy for prolate and oblate shapes. 
This feature which has been illustrated by few isolated  prolate ($^{76}$Ge, $^{152}$Sm, $^{240}$Pu) and oblate ($^{196}$Pt) nuclei 
in Fig. \ref{inter}, is now investigated more systematically. 
For this purpose, we plot in Fig. \ref{defsys} the centroid energies for each $K$ angular momentum projection for all the deformed even-even nuclei considered in this work. 
As can be observed the centroid energies of the $K^\pi=0^-$ components are systematically lower than those of the  $K^\pi=1^-$ components for prolate nuclei and the other way round for oblate nuclei.  
Above the $A$=150 mass region, the large difference of energy between $K^\pi=0^-$ and $K^\pi=1^-$ centroids is related to the large intrinsic deformation of the rare earth nuclei (see for example Fig. 1 of Ref. \cite{hj_epja_2007}). 
In contrast in the $A$=130 mass region, the centroid energies of the $K^\pi=0^-$ and $K^\pi=1^-$ components are close to each other.  In such a situation, the experimental strength might not show the typical double hump pattern usually associated with a deformed nucleus. 
Let us notice that similar $K$ hierarchy has also been obtained for quadrupole resonances \cite{PG08,SL14}. 

\subsection{Comparison with experimental photoabsorption data} \label{rawcompexp}

The axially-symmetric-deformed HFB+QRPA method is now applied 
to provide $\gamma$-ray strength function for all nuclei for which photoabsorption data have been analyzed and compiled in the IAEA RIPL library \cite{ripl3}. In this compilation, a large part of the nuclei are not even-even, so that they can not be, strictly speaking, compared with our theoretical predictions limited here to even-even nuclei. 
For odd-mass (resp odd-odd) nuclei, experimental data are associated with theoretical calculations performed for the 2 (resp. 4) neighbouring even-even nuclei. The analysis in the RIPL library consisted in fitting experimental photoabsorption data with one Lorentzian function or two if necessary. We therefore use two methods to analyze our predictions. When a single Lorentzian is recommended in RIPL, we compare its peak energy with the centroid of the total strength, obtained by summing the $K^\pi=0^-$ and $|K|^\pi=1^-$ components if the nucleus is theoretically predicted to be deformed. When two Lorentzian functions are recommended in RIPL, we compare the energy of the lowest (resp.  highest) peak with the centroid of the $K^\pi=0^-$ (resp. $|K|^\pi=1^-$) component for nuclei theoretically predicted to be prolate and with the centroid of the $|K|^\pi=1^-$ (resp. $K^\pi=0^-$) component for nuclei theoretically predicted to be oblate, following the conclusions reached in Sect. \ref{sbsec_def}. 
(Keeping in mind that for HFB predicted spherical nuclei the $K$ centroids are degenerated).  

\begin{figure}[h]
\begin{center}
\includegraphics[trim=1cm 11cm 0cm 0cm, clip=true, scale=0.6]{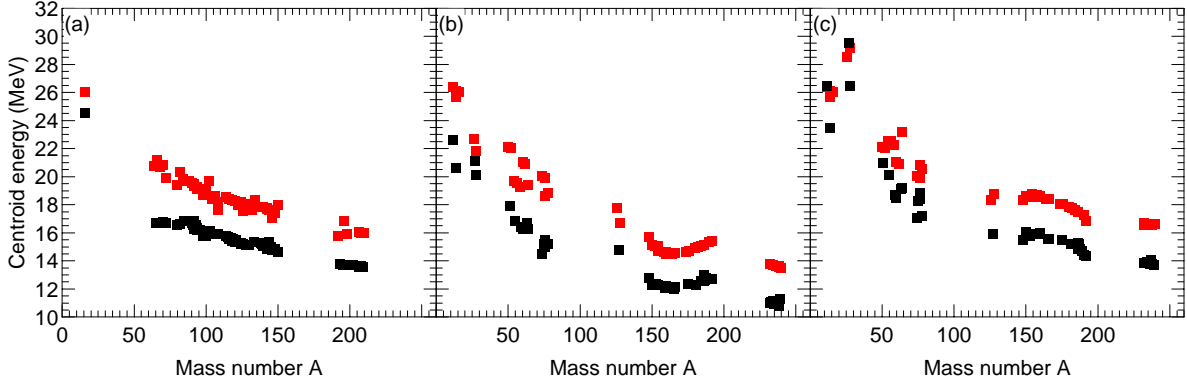}
\caption{Comparison of the QRPA centroid energies (red) with the peak energies compiled in the RIPL library (black). (a) Nuclei for which a single Lorentzian is recommended. (b,c) Lowest (b) and highest (c) peak energies for the nuclei for which two Lorentzians are recommended.}
\label{D1Mexp}
\end{center}
\end{figure}

The comparisons are shown in Fig. \ref{D1Mexp}. 
As can be observed, our predictions display a qualitatively good  mass dependence compared to experimental data, but with a systematic overestimate of the order of 2 MeV. As already mentioned in Sect. \ref{sec_mod}, this shift already observed in Ref.~\cite{Per11} is expected since our QRPA calculations do not account for the effects related to $qp$ configurations involving more than 2-$qp$ states or to the phonon coupling. Nevertheless, the mass dependence of the centroid energy follows particularly well the experimental trend, a feature particularly striking in the rare earth region. 
This systematic overestimation can be removed by a global ($\sim$2 MeV) shift of the states involved in the response. 
Furthermore a systematic $\sim$2 MeV downward energy shift should reduce the D1M EWSRs (around $\sim$1.5 TRK units) to reach the experimental values which rarely exceed 1.3 TRK units.  

\subsection{Models for folded strength functions} \label{sect_phon}

The QRPA provides a rather satisfactory description of the GDR centroid and the fraction of the EWSR exhausted by the $E1$ mode. 
However the GDR is known experimentally to have a relatively large width and therefore a finite lifetime and  
for a proper description of experimental data, it is necessary to go beyond the QRPA scheme. 
Different microscopic theories (such as second RPA and particle-phonon coupling) explain the location and width of the GDR. 
For the sake of simplicity and applicability to a large number of nuclei of astrophysical interest, we restrict ourselves to the semi-empirical broadening of the GDR already introduced in Sec.~\ref{formalism_4qp}. 
Such a broadening is obtained
by folding the QRPA $B(E1)$ distribution
by a normalized Lorentzian function, as given by Eq.~(\ref{broad}) and Eq.~(\ref{eq_folding})
where both the width $\Gamma$  and energy shift $\Delta$ can be adjusted on experimental data. Due to the phenomenological character of such an approach, three different prescriptions are adopted to estimate $\Delta$ and $\Gamma$. In the first model (hereafter referred to as Model 0), both parameters are assumed to be independent of the energy and identical for all nuclei. More precisely, the values of $\Delta=2$~MeV and $\Gamma=2.5$~MeV are chosen to globally reproduce experimental photoabsorption data. For the other two models, another strategy is adopted, the energy shift and the width $\Gamma$ is adjusted on each photoabsorption cross section available experimentally; 
in addition,  the energy shift is this time taken energy-dependent following the simple relation $\Delta(\omega)=\Delta_0+\Delta_{4qp}(\omega)$, where $\Delta_0$ is a constant shift due to the coupling between $qp$-states and phonons and the quantity $\Delta_{4qp}(\omega)$ is an extra shift which empirically describes the effect of complex configurations and for this reason is taken to be proportional to the number of 4-$qp$ states,  $n^K_{4qp}(\omega)$, as defined  in Sect.~\ref{formalism}. The latter correction therefore varies with the excitation energy $\omega$ and obviously depends on the nucleus considered. Two different prescriptions are considered to estimate this extra energy shift. The first approximation (hereafter referred to as Model 1) assumes that $\Delta^K_{4qp}=\delta_{4qp} \times n^K_{4qp}(\omega)/n^K_{4qp}(\omega=30{\rm MeV})$ where $\delta_{4qp}$ is a parameter adjusted on experimental photoabsorption cross sections. In this model, $\Delta^K_{4qp}$ is arbitrarily normalized  to the $n^K_{4qp}$ value at $\omega=30$~MeV. The second approximation (Model 2) takes into account the number of 4-$qp$ states relative to the number of 2-$qp$ states at the excitation energy $\omega$ and reads $\Delta^K_{4qp}=\delta_{4qp} \times n^K_{4qp}(\omega)/n^K_{2qp}(\omega)$.

\begin{figure}
\begin{center}
\includegraphics[scale=0.75]{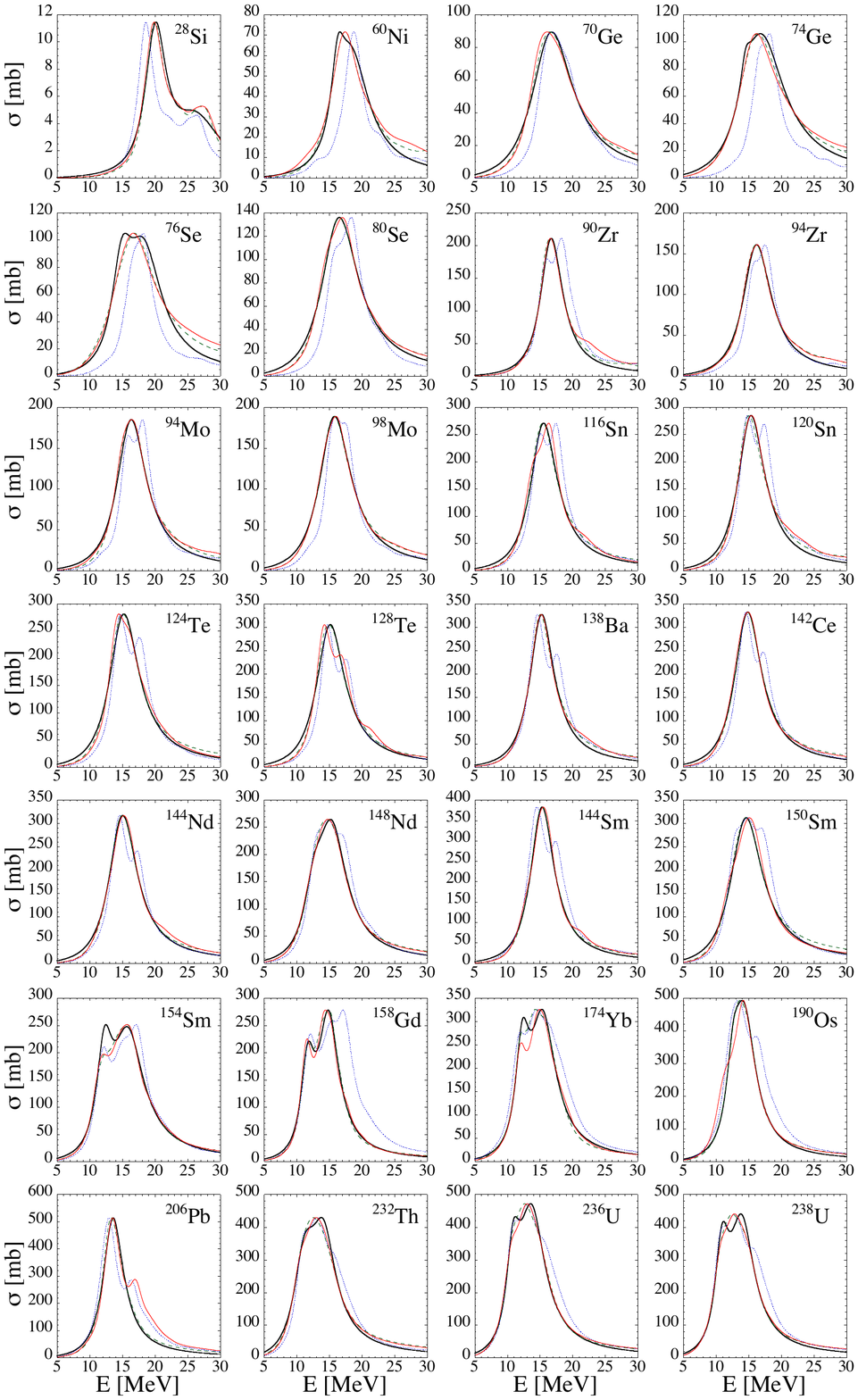}  
\caption{Comparison for 28 nuclei between the experimental photoabsorption cross section \cite{ripl3} (black solid line) and the fits corresponding to Model 0 (dotted blue line), Model 1 (green dashed line) and Model 2 (red solid line). The experimental cross section around the GDR is represented by one or two parametrized Lorentzian functions, as traditionally done \cite{ripl3}.}
\label{fig_xs_all}
\end{center}
\end{figure}

\begin{figure}
\begin{center}
\includegraphics[scale=0.5]{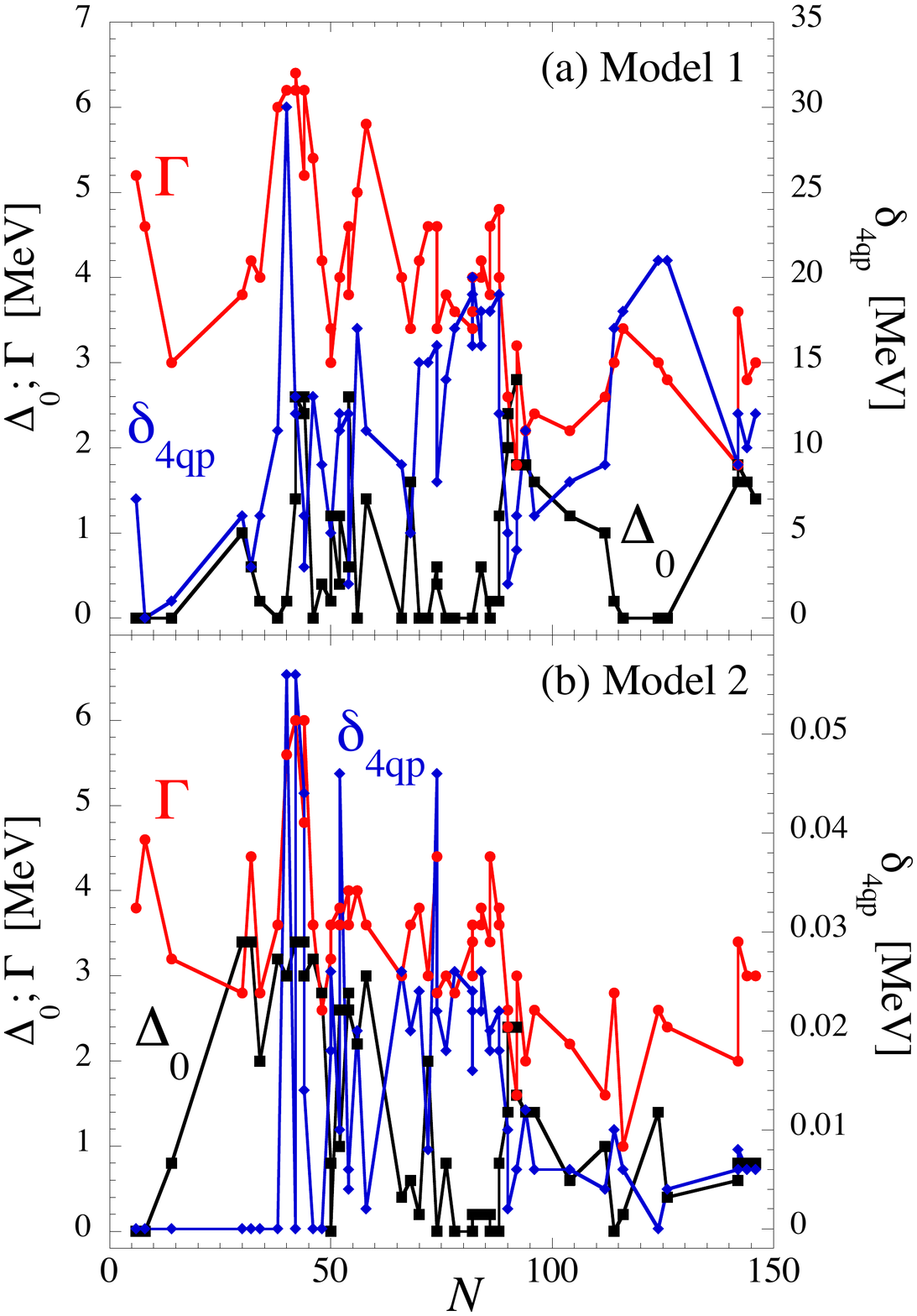}  
\caption{Values of the three parameters $\Delta_0$ (squares), $\Gamma$ (circles) and $\delta_{4qp}$ (diamonds) adjusted on experimental photoabsorption cross sections as a function of the neutron number of the corresponding nucleus for Models 1 (upper panel) and 2 (lower panel).}
\label{fig_param}
\end{center}
\end{figure}

In Fig. \ref{fig_xs_all}, QRPA photoabsorption cross sections are compared with experimental data \cite{ripl3}  for a sample of 28 even-even nuclei. Note that experimental cross sections are represented by parametrized Lorentzian functions in the vicinity of the GDR, i.e. cross sections at the energies significantly lower or larger than the GDR centroid should not be considered. As seen in Fig. \ref{fig_xs_all}, constant energy shift and width (Model 0) globally reproduce experimental data, in particular the low-energy tail of the cross section, though the width tends to be too low for light and medium-$A$ nuclei. A 
split peak above the GDR is also present in the QRPA strength of some spherical nuclei, though not observed experimentally. 
As stressed for example in Refs. \cite{Sarchi:2004pf,Peru:2005di} it is expected that the fragmentation is somewhat reduced by the coupling of the RPA modes to 2p-2h states and equivalently by the coupling of the QRPA modes to 4-$qp$ states. 
The introduction of an energy-dependent shift proportional to the number of 4-$qp$ states is seen to cure this deficient pattern, the second peak being shifted into the low energy one. For this reason, Models 1 and 2 much better describe experimental data than Model 0. Since each parameter has been adjusted for each nucleus to optimize the GDR properties, excellent agreement could be obtained for both spherical and deformed nuclei in most cases. The corresponding parameters $\Delta_0$, $\Gamma$ and $\delta_{4qp}$ adjusted for each photoabsorption cross section available experimentally for even-even nuclei are plotted in Fig.~\ref{fig_param}. 

In addition to the phenomenological corrections, questions still arise i) for experimentally unknown nuclei for which values of the $\Delta_0$, $\Gamma$ and $\delta_{4qp}$ parameters need to be assigned  and ii) for the odd-$A$ and odd-odd nuclei for which no HFB+QRPA calculations can be consistently performed nowadays with the same accuracy. As it appears in Fig.~\ref{fig_param} the adjustment of the $\Delta_0$, $\Gamma$ and $\delta_{4qp}$ parameters is clearly nucleus dependent without any evident $N$ or $A$ behavior. 
Furthermore the choice of the normalization of the $4qp$ state densities ($n^K_{4qp}(\omega)/n^K_{4qp}(\omega=30{\rm MeV})$ for Model 1 and 
$n^K_{4qp}(\omega)/n^K_{2qp}(\omega)$ for Model 2) leads to different constant shift values $\Delta_0$, 
reflecting the limitation of the present strategy in the introduction of ingredients going beyond QRPA formalisms. It also allows us to test the sensitivity of our predictions with respect to different prescriptions for this phenomenological procedure.   
For experimentally unknown nuclei, the use of a smooth function fitting globally the $N$ or $A$ dependencies would inevitably give rise to rather larger uncertainties. Then an interpolation procedure is followed to estimate in Models 1 and 2 the parameter values when they cannot be tuned on experimental GDR data. Since the GDR width is known to be shell dependent \cite{go98}, hence essentially a function of the neutron number, the interpolation is performed as a function of the neutron number $N$. 

\begin{figure}
\begin{center}
\includegraphics[scale=0.4]{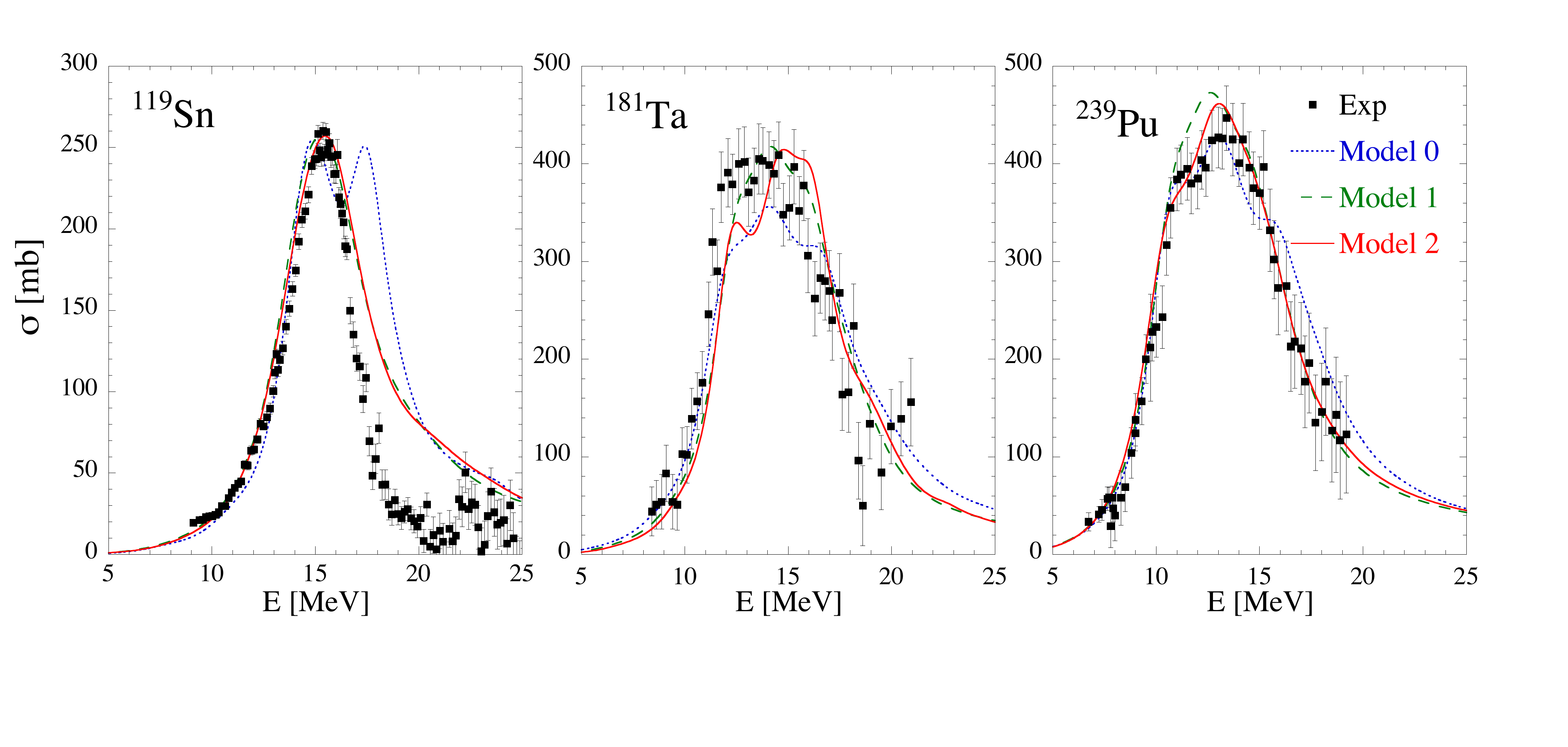}  
\vskip -1. cm
\caption{Comparison between experimental photoabsorption (or photoneutron) cross sections \cite{fultz69,gur81,demoraes93,gur76} (squares) and the QRPA ones obtained from the interpolated strengths corresponding to Model 0 (dotted line), Model 1 (dashed line) and Model 2 (solid line) for 3 odd-A nuclei.}
\label{fig_odd}
\end{center}
\end{figure}

Concerning the odd-A and odd-odd nuclei, the $B(E1)$ distributions have not been calculated yet. To estimate their strength $S_{E1}$, we have designed an interpolation procedure using the QRPA predictions of the even-even nuclei based on the relative smooth variation of GDR between neighboring nuclei \cite{iaea00}.
The procedure consists in determining the energy dependence of the strength of an odd nucleus from a geometrical mean of the neighboring even-even nuclei and normalizing the strength to the interpolated EWSR. For example, for an odd neutron number $N$ in a even $Z$ chain: $S_{E1}(Z,N)\propto \sqrt{S_{E1}(Z,N-1) \times S_{E1}(Z,N+1)}$. This procedure has been tested and found to give satisfactory results as shown in Fig.~\ref{fig_odd} for  three odd-$A$ nuclei.


\subsection{Low-lying strength}
\begin{figure}
\begin{center}
\includegraphics[scale=0.4]{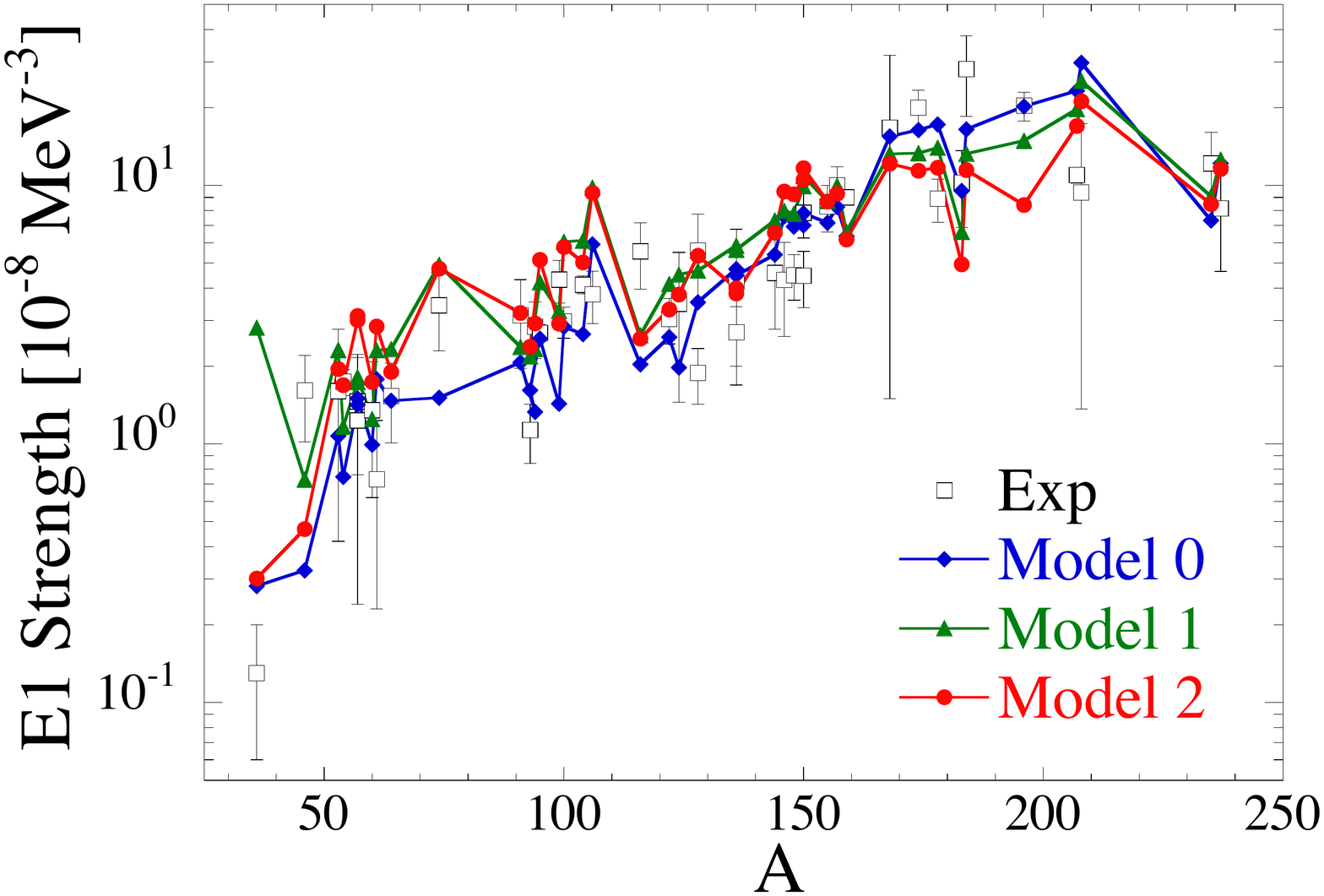}  
\caption{Comparison of the QRPA low-energy $E1$-strength functions
with the experimental compilation
\cite{ripl3} including resolved-resonance and thermal-captures measurements, as well as
photonuclear data for nuclei from $^{36}{\rm Cl}$ up to $^{239}{\rm U}$ at energies
ranging from 4 to 8 MeV. }
\label{fig_fe1_exp}
\end{center}
\end{figure}

Up to now we have discussed the QRPA results focusing on the GDR region.
It is well known that the knowledge of the low-energy part of the $\gamma$ strength is crucial for the photo-nuclear reactions description. 
This part of the energy spectrum is also provided by our QRPA approach.  
In Fig.~\ref{fig_fe1_exp} the predictions are compared  with the compilation of experimental $E1$ strength functions at energies
ranging from 4 to 8 MeV
\cite{ripl3} for nuclei from $^{36}{\rm Cl}$ up to $^{239}{\rm U}$. The data set
includes resolved-resonance measurements, thermal-captures measurements and photo-nuclear
data. In some cases the original experimental values need to be corrected,
typically for non-statistical effects, so that only values
recommended in Ref.~\cite{ripl3} are considered in Fig.~\ref{fig_fe1_exp}. 
The three models considered here reproduce well the trend of the strength with respect to the mass number and globally agree, within the error bars, with the low-energy E1 strength. 
The rms deviation on the 44 theoretical to experimental ratios amounts to $f_{\rm rms}=1.78, 2.02$ and 1.84 for Models 0, 1 and 2, respectively. 
This degree of accuracy is similar to the one found with the previous Skyrme-HFB plus QRPA calculation \cite{Gor02,Gor04}.

\subsection{Application to exotic neutron-rich nuclei}
\label{sec_nrich}
\begin{figure}
\begin{center}
  \includegraphics[width=0.9\columnwidth]{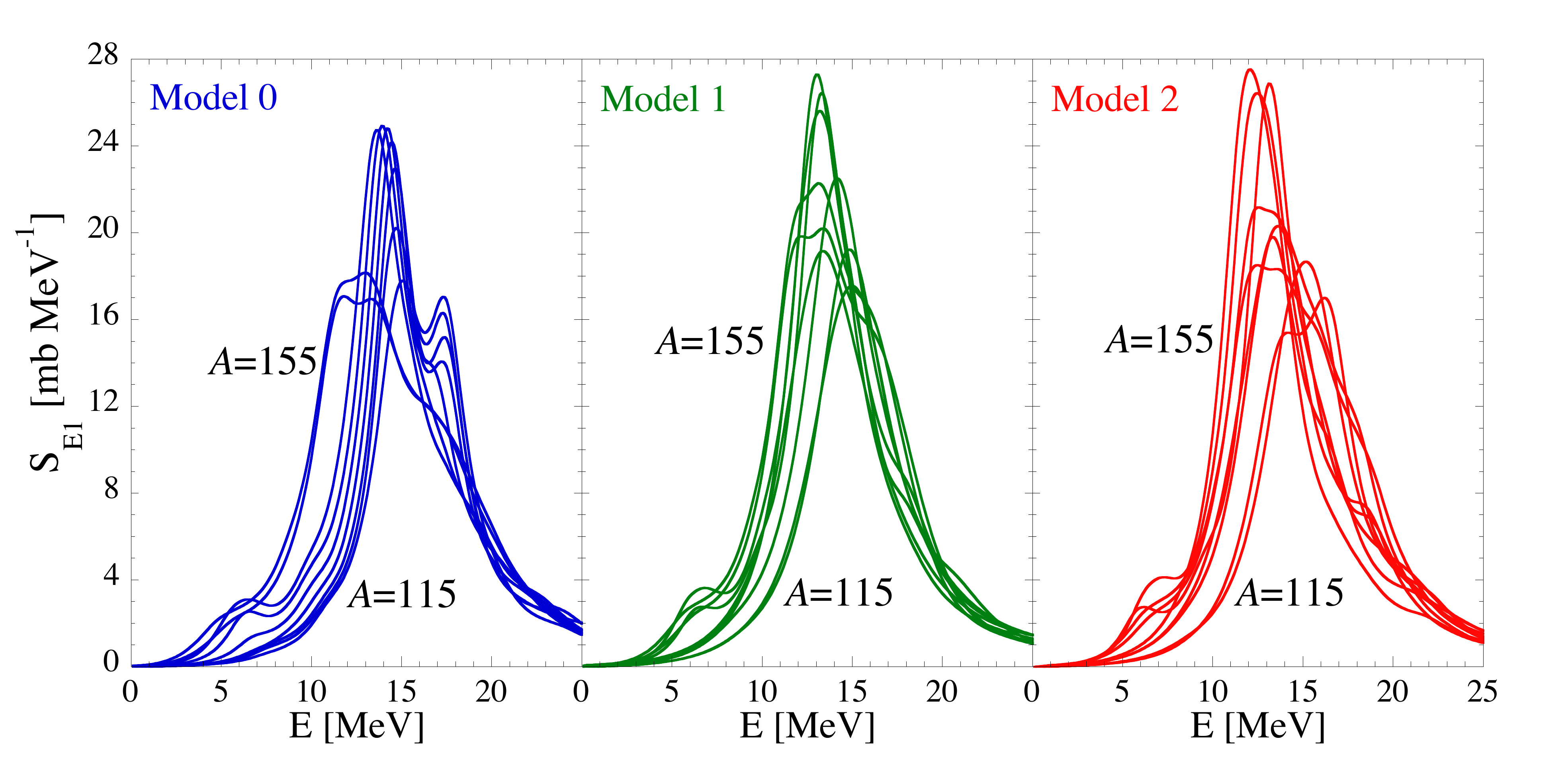}  
  \caption{Comparison between the $E1$ strength functions for Sn isotopes (from $A=115$ to 155 by steps of $\Delta A=5$) obtained with 
the three prescriptions used to correct the HFB+QRPA model based on the D1M force.}
\label{fig_fE1_D1M_Sn}
\end{center}
\end{figure}

\begin{figure}
\begin{center}
  \includegraphics[width=0.9\columnwidth]{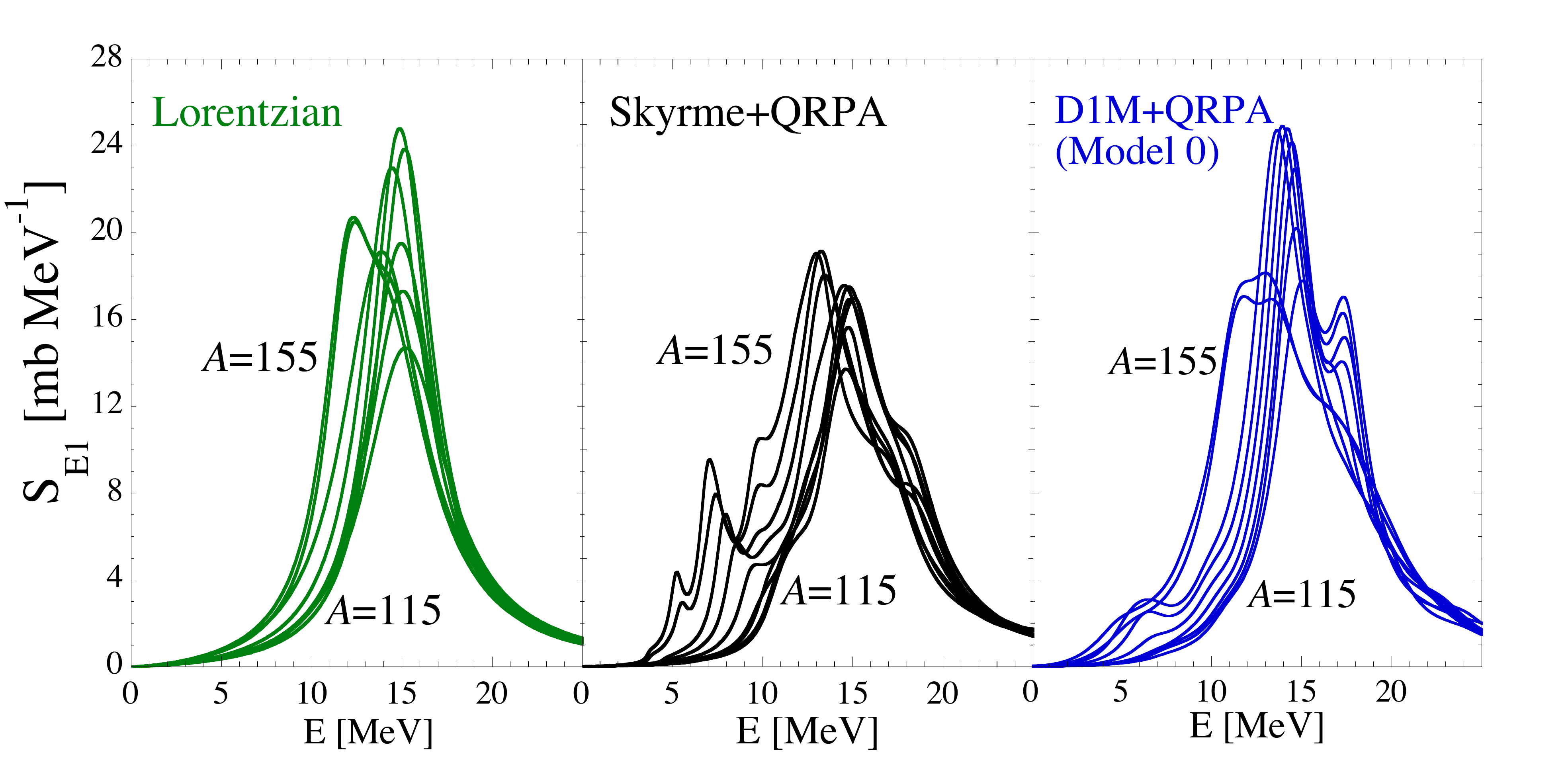}  
  \caption{Comparison between the $E1$ strength functions for Sn isotopes (from $A=115$ to 155 by steps of $\Delta A=5$) obtained with 
the Lorentzian model \cite{ripl3} (left panel), 
 the microscopic QRPA calculation based on the Skyrme 
force \cite{Gor04} (central panel) and the present QRPA calculation with the D1M Gogny force (Model 0) (right panel).}
\label{fig_fE1_Sn}
\end{center}
\end{figure}

The theoretical approach presented here is based on microscopic QRPA calculations but it contains some phenomenological ingredients needed to reproduce reaction observables. The question of the predictive power of such an approach naturally arises when dealing with exotic nuclei. 
Figure~\ref{fig_fE1_D1M_Sn} shows the  $E1$ $\gamma$-ray strength functions for $^{115-155}$Sn  isotopes obtained with our three models (see Sec. \ref{sect_phon}). 
Although the GDR predictions could be different, the appearance of some extra strength around 5~MeV for Sn isotopes above the $N=82$ shell closure is independent of the folding prescriptions.
This low-lying strength is however quite different from the one predicted by other models, as shown in Fig.~\ref{fig_fE1_Sn} where the 
Generalized Lorentzian (GLO) \cite{ripl3} and the Skyrme-HFB+QRPA \cite{Gor04} results are compared to the D1M-HFB+QRPA Model 0.
For such exotic neutron-rich nuclei, the Skyrme-HFB plus QRPA calculation \cite{Gor04} predicts pygmy resonances more enhanced than the Gogny ones, while the GLO approach cannot predict by definition any additional strength at low energy. Nevertheless for nuclei close to the valley of $\beta$-stability the
microscopic $E1$ strength functions  look rather similar to the phenomenological Lorentzian.

\begin{figure}
\begin{center}
  \includegraphics[width=0.9\columnwidth]{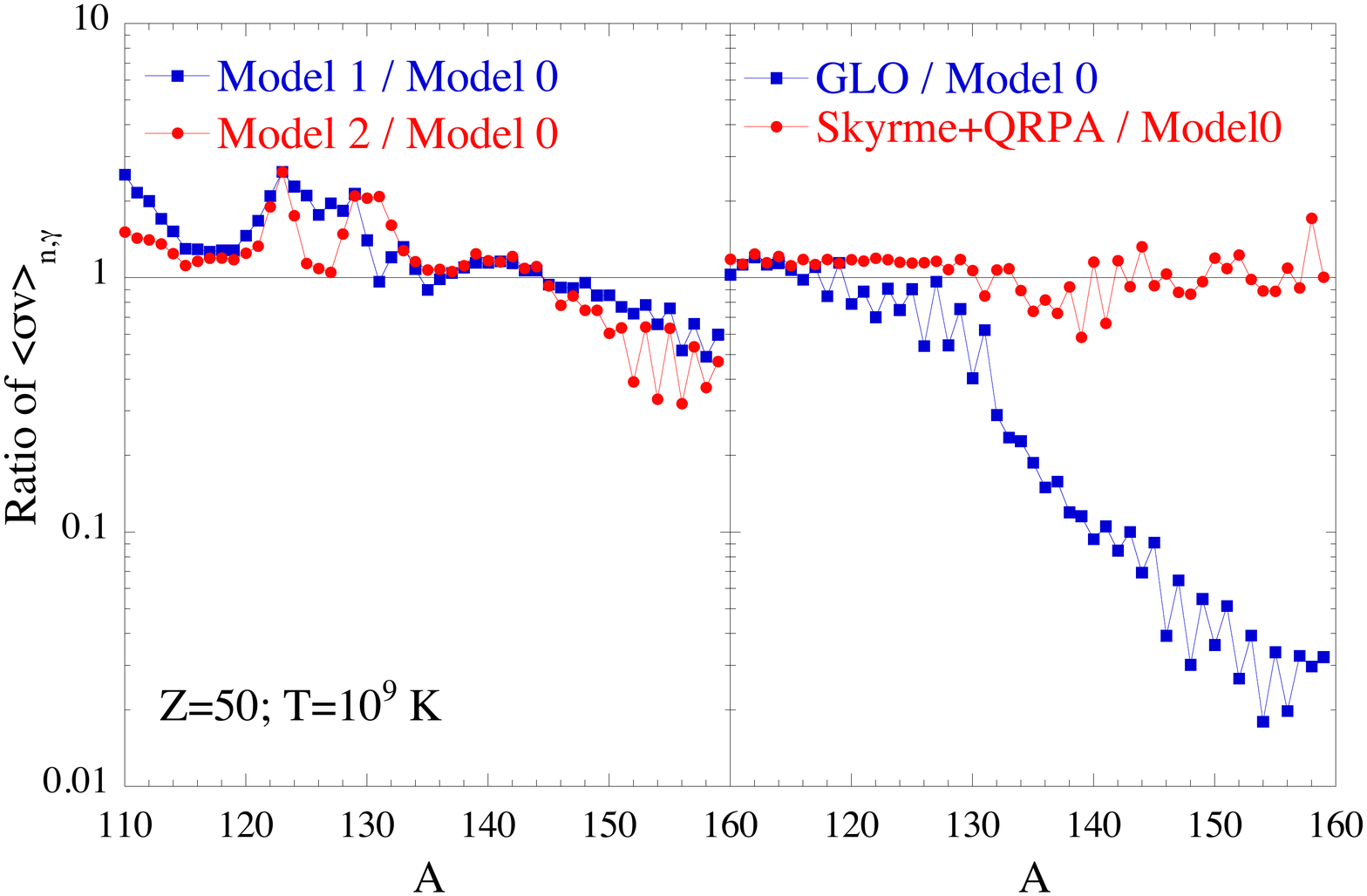}  
  \caption{Left panel: Ratio of the Maxwellian-averaged radiative neutron capture rate (at a temperature of 10$^9$ K) for the Sn isotopes obtained with the three present D1M+QRPA models; Right panel: same where the neutron capture rate with Model 0 of the D1M+QRPA strength is compared with the one obtained with 
the GLO model (squares) \cite{ripl3} and to the one predicted with the Skyrme-HFB+QRPA model based on the BSk7 force  (circles) \cite{Gor04}.}
\label{fig_ng_Sn}
\end{center}
\end{figure}

The differences illustrated in Figs.~\ref{fig_fE1_D1M_Sn}-\ref{fig_fE1_Sn} between the various $\gamma$-ray strength predictions for nuclei far from the valley of stability may also have a significant impact on the predicted neutron-capture cross section of astrophysical interest.
In order to investigate this point, the Maxwellian-averaged neutron capture rate are calculated with the TALYS code \cite{Hil11,Gor08,Kon08,Kon12} using the
$\gamma$-ray strength obtained with the three prescribed D1M+QRPA models as well as with the GLO \cite{ripl3} and  Skyrme-HFB+QRPA modelsl \cite{Gor04}. Note that, for consistency, in the present calculation of the radiative neutron capture rates, the D1M masses \cite{Gor09} and the HFB plus combinatorial nuclear level densities based on the D1M single-particle and pairing properties \cite{Hil12} are adopted. 
Figure \ref{fig_ng_Sn} shows the Maxwellian-averaged neutron capture rate of the Sn isotopes for Models 1 and 2 (left panel) and GLO and Skyrme-HFB+QRPA (right panel) with respect to the Model 0 predictions  used as a reference. As can be seen on the left panel of Fig. \ref{fig_ng_Sn}, the three prescriptions used to fold the original D1M+QRPA strength predict radiative neutron capture rates within a factor of 2, even for the most exotic nuclei. 
Similarly, the reaction rates obtained using the D1M+QRPA strengths agree rather well with those based on the Skyrme-HFB plus QRPA along the whole isotopic chain. This similarity can be explained by the rather same strength predicted at low energies below the GDR, although the Gogny calculation predicts a strength spread on a much wider energy range.
However, for exotic neutron-rich Sn isotopes our present QRPA $\gamma$-ray strength is seen to give rise to reaction rates that can be about 30 times larger than those obtained with the GLO model. 

\section{Conclusions}
\label{sec_conc}

Large scale calculations of $E1$ $\gamma$-ray strength functions for even-even nuclei have been undertaken 
within the axially-symmetric-deformed HFB+QRPA approach in a fully consistent way using the finite-range Gogny interaction. 
The convergence of the numerical calculation has been analyzed with respect to the size of the basis (number of HO shells) 
and the 2-$qp$ excitation energy. This analysis allowed to establish practical choices for large scale calculations optimizing both the minimization of the computational cost and the convergence. Predictions obtained for two parameter sets of the Gogny interaction, namely D1S and D1M, have been compared and shown to give rise to a similar global behaviour. Nevertheless, the D1M $E1$ strength is found to be systematically shifted, leading to lower centroid energy  and a smaller EWSR in comparison with D1S. 

The role of the intrinsic deformation has been systematically investigated. The split between $K$=0 and $|K|$=1 total angular momentum projections for deformed nuclei as well as their opposite hierarchy for prolate and oblate shapes has been confirmed in the whole nuclear chart.  
Large deformations give rise to a double peak structure while small deformations lead to a split too small to be disentangled. 

We have calculated the $E1$ $\gamma$-ray strength for all even-even nuclei for which photoabsorption data exist. 
The comparison between QRPA and measured cross sections revealed a systematic energy shift of about $\sim$2 MeV of the Gogny-QRPA strength with respect to experimental data. 
This energy shift (of $\sim$2 MeV) induces that the theoretical values of the EWSR obtained for all the nuclei studied in the present 
work are systematically larger than the experimental ones. 
Three prescriptions have been proposed to cure this discrepancy. They correspond to a folding procedure of the $B(E1)$ discrete distribution on the basis of a Lorentzian function, which not only shifts the $E1$ strength down by more or less 2~MeV but also widen the distribution, as experimentally observed in photoabsorption cross sections.
Such a folding is also needed to produce $S_{E1}$ strength functions as inputs in reaction models (\textit{e.g.} TALYS code). 
Two of these prescriptions fit the experimental photoabsorption data by adjusting the width and energy shift parameters of the Lorentzian, taking into account the density of dipole 4-$qp$ excitations. 
This microscopic ingredient allows us to investigate effects beyond the standard QRPA description involving only 2-$qp$ excitations.  The three prescriptions are found to give a satisfactory description of experimental data and are shown to provide globally similar $E1$ strength. 
In particular the neutron capture rates of astrophysical interest do not differ by more than a factor of 2 for the three prescriptions, even far away from the valley of $\beta$-stability.

Further improvements of the present approach can be envisioned. First, the interpolation procedure for odd nuclei can be replaced by a fully microscopic QRPA calculation of odd systems. Second, the folding procedure can be improved by including microscopically the particle-phonon coupling. Third, dynamic deformations can be considered when differing from the HFB one (see \textit{e.g.} Fig. 3 of Ref. \cite{Delaroche:2009fa}). 
The present encouraging results, in parallel to those obtained for nuclear masses \cite{Gor09} and  nuclear level densities \cite{Hil12}, will allow us to include such microscopic ingredients (obtained on the basis of one unique Gogny interaction) into cross sections calculations. 
We believe that working along such a path is a way, in the future, to improve cross section evaluations and predictions on the basis of reliable and accurate microscopic inputs.

\section*{Acknowledgments}
We acknowledge PRACE for awarding us access to resource CURIE based in FRANCE at TGCC-CEA. SG acknowledges the support of the FRS-FNRS and 
MM of the ``Espace de Structure et de r\'eactions Nucl\'eaire Th\'eorique'' (ESNT, \url{http://esnt.cea.fr} ) at CEA.
We also wish to thank Isabelle Deloncle for stimulating discussions and her constructive comments.



\begin{thebibliography}{9}
\bibitem{arnould07} 
  M.~Arnould, S.~Goriely and K.~Takahashi,
  Phys.\ Rept.\  {\bf 450}, 97 (2007).




\bibitem{go98} S. Goriely S,  Phys. Lett. B {\bf 436}, 10 (1998).
\bibitem{xu14} Y. Xu, S. Goriely, A.J. Koning, S. Hilaire, Phys. Rev. C {\bf 90}, 024604 (2014).
\bibitem{ripl3} R. Capote {\it et al.} {\it Nucl. Data Sheets} {\bf 110} 3107 (2009).

\bibitem{Gor02} 
  S.~Goriely and E.~Khan,
  Nucl.\ Phys.\ A {\bf 706}, 217 (2002).
  
\bibitem{Gor04} 
  S.~Goriely, E.~Khan and M.~Samyn,
  Nucl.\ Phys.\ A {\bf 739}, 331 (2004).






\bibitem{PG08} 
  S.~P\'eru and H.~Goutte,
  Phys.\ Rev.\ C {\bf 77}, 044313 (2008).

\bibitem{Mar11} 
  M.~Martini, S.~P\'eru and M.~Dupuis,
  Phys.\ Rev.\ C {\bf 83}, 034309 (2011).


\bibitem{Per11} 
  S.~P\'eru, G.~Gosselin, M.~Martini, M.~Dupuis, S.~Hilaire and J.-C.~Devaux,
  Phys.\ Rev.\ C {\bf 83}, 014314 (2011).






\bibitem{Martini:2014ura} 
  M.~Martini, S.~P\'eru and S.~Goriely,
  Phys.\ Rev.\ C {\bf 89}, no. 4, 044306 (2014).

\bibitem{DechargeNPA83}
J.~Decharge and L.~Sips, Nucl.\ Phys.\ A {\bf 407}, 1 (1983).  

\bibitem{Per14} 
  S.~P\'eru and M.~Martini,
  Eur.\ Phys.\ J.\ A {\bf 50}, 88 (2014).
  



\bibitem{book_ring_schuck}
P. Ring and P. Schuck. 1980. Springer. The Nuclear Many-Body Problem (N.Y.).

  
\bibitem{Yannouleas:1983kp} 
  C.~Yannouleas, M.~Dworzecka and J.~J.~Griffin,
  Nucl.\ Phys.\ A {\bf 397}, 239 (1983).
  
  
\bibitem{Yannouleas:1987zz} 
  C.~Yannouleas,
  Phys.\ Rev.\ C {\bf 35}, 1159 (1987).
  
\bibitem{Bertsch:1983zz} 
  G.~F.~Bertsch, P.~F.~Bortignon and R.~A.~Broglia,
  Rev.\ Mod.\ Phys.\  {\bf 55}, 287 (1983).

\bibitem{Sol76} V.G. Soloviev, {\it Theory of complex nuclei} (Oxford: Pergamon Press, 1976). 
\bibitem{Sol78} V.G. Soloviev, Ch. Stoyanov and V.V. Voronov, Nucl. Phys.\textbf{ A304}, 503 (1978).
\bibitem{Tso08} N. Tsoneva and H. Lenske, Phys. Rev. C {\bf 77}, 024321 (2008).  

\bibitem{pap09} 
  P.~Papakonstantinou and R.~Roth,
  Phys.\ Lett.\ B {\bf 671}, 356 (2009).


\bibitem{gam11} 
  D.~Gambacurta, M.~Grasso and F.~Catara,
  Phys.\ Rev.\ C {\bf 84}, 034301 (2011).



\bibitem{gam12} 
  D.~Gambacurta, M.~Grasso, V.~De Donno, G.~Co' and F.~Catara,
  Phys.\ Rev.\ C {\bf 86}, 021304 (2012).




\bibitem{Gambacurta:2015pva} 
  D.~Gambacurta, F.~Catara, M.~Grasso, M.~Sambataro, M.~V.~Andres and E.~G.~Lanza,
  Phys.\ Rev.\ C {\bf 93}, no. 2, 024309 (2016).


\bibitem{Colo:2001fz} 
  G.~Colo and P.~F.~Bortignon,
  Nucl.\ Phys.\ A {\bf 696}, 427 (2001).


\bibitem{kame04} S. Kamerdzhiev, J. Speth, G. Tertychny, Phys. Rep. \textbf{393}, 1 (2004).

\bibitem{Sarchi:2004pf} 
  D.~Sarchi, P. F. Bortignon, G. Colo
  Phys.\ Lett.\ B {\bf 601}, 27 (2004).


\bibitem{avde11} A. Avdeenkov, S. Goriely, S. Kamerdzhiev, S. Krewald,  Phys. Rev. C {\bf 83}, 064316 (2011).
\bibitem{acha15} O. Achakovskiy, A. Avdeenkov, S. Goriely, S. Kamerdzhiev, S. Krewald,  Phys. Rev. C {\bf 91}, 034620 (2015).


\bibitem{Gor09} 
  S.~Goriely, S.~Hilaire, M.~Girod and S.~P\'eru,
  Phys.\ Rev.\ Lett.\  {\bf 102}, 242501 (2009).






\bibitem{Berman:1975tt} 
  B.~L.~Berman and S.~C.~Fultz,
  Rev.\ Mod.\ Phys.\  {\bf 47}, 713 (1975).

\bibitem{SL14}  G.~Scamps and D.~Lacroix,  Phys.\ Rev.\ C {\bf 89}, 034314 (2014).

\bibitem{VanIsacker:1992zz} 
  P.~Van Isacker, M.~A.~Nagarajan and D.~D.~Warner,
  Phys.\ Rev.\ C {\bf 45}, R13 (1992).










\bibitem{Yoshida:2008rw} 
  K.~Yoshida and N.~Van Giai,
  Phys.\ Rev.\ C {\bf 78}, 014305 (2008).
  
\bibitem{Kleinig:2008gq} 
  W.~Kleinig, V.~O.~Nesterenko, J.~Kvasil, P.-G.~Reinhard and P.~Vesely,
  Phys.\ Rev.\ C {\bf 78}, 044313 (2008).
  
  
\bibitem{Arteaga:2009mb} 
  D.~P.~Arteaga and P.~Ring,
  Phys.\ Rev.\ C {\bf 77}, 034317 (2008).


\bibitem{Arteaga:2008ej} 
  D.~P.~Arteaga, E.~Khan and P.~Ring,
  Phys.\ Rev.\ C {\bf 79}, 034311 (2009).



  
\bibitem{Losa:2010bm} 
  C.~Losa, A.~Pastore, T.~Dossing, E.~Vigezzi and R.~A.~Broglia,
  Phys.\ Rev.\ C {\bf 81}, 064307 (2010).
  





  
\bibitem{hj_epja_2007}
S. Hilaire, M. Girod, 
Eur.\ Phys.\ J.\ A {\bf 33}, 237 (2007).

\bibitem{Peru:2005di} 
  S.~Peru, J.~F.~Berger and P.~F.~Bortignon,
  Eur.\ Phys.\ J.\ A {\bf 26}, 25 (2005).

\bibitem{iaea00} Photonuclear data for applications; cross sections and spectra, IAEA-Tecdoc-1178 (2000).

\bibitem{fultz69} S.C. Fultz, B.L. Berman, J.T. Caldwell, R.L. Bramblett,  M.A. Kelly, Phys. Rev. {\bf 186}, 1255 (1969).
\bibitem{gur81} G.M. Gurevich, L.E. Lazareva, V.M. Mazur, S.Yu. Merkulov, G.V. Solodukhov, V.A. Tyutin,  Nucl. Phys. {\bf A351},  257 (1981).
\bibitem{demoraes93} M.A.P.V. De Moraes, M.F. Cesar, Physica Scripta {\bf 47}, 519 (1993).
\bibitem{gur76} G.M. Gurevich, L.E. Lazareva, V.M. Mazur, G.V. Solodukhov, B.A. Tulupov, Nucl. Phys. {\bf A273} 326 (1976).

\bibitem{Hil11} S. Hilaire, A. J. Koning and S. Goriely, Journal of the Korean Physical Society {\bf 59}, 767 (2011).
\bibitem{Gor08} S. Goriely, S. Hilaire and A. J. Koning, Astronomy and Astrophysics {\bf 487},  767 (2008).
\bibitem{Kon08} A. J. Koning, S. Hilaire and M. C. Duijvestijn  Nucl. Data for Science and technology (EDP Sciences, eds Bersillon et al.) 211 (2008).
\bibitem{Kon12} A.J. Koning and D. Rochman, Nuclear Data Sheets {\bf 113}, 2841 (2012).
\bibitem{Hil12} S. Hilaire, M. Girod, S. Goriely S and A. J. Koning, Phys. Rev. C {\bf 86}, 064317 (2012).


\bibitem{Delaroche:2009fa} 
  J.~P.~Delaroche, M.~Girod, J.~Libert, H.~Goutte, S.~Hilaire, S.~P\'eru, N.~Pillet and G.~F.~Bertsch,
  Phys.\ Rev.\ C {\bf 81}, 014303 (2010).




\end{thebibliography}
\end{document}